\documentclass[aps, 12pt, preprint, preprintnumbers, amsmath,amssymb,nofootinbib,superscriptaddress,hyperref]{revtex4-1} 
\usepackage{graphicx}
\usepackage{slashed}
\usepackage{rotating}
\usepackage{amsmath,amssymb,color,colortbl}
\usepackage[right=28truemm,left=28truemm,top=30truemm,bottom=30truemm]{geometry}
\usepackage{fancyhdr}
\usepackage{here}
\usepackage[subrefformat=parens]{subcaption}
\usepackage[samesize]{cancel} 
\usepackage{comment}
\usepackage{caption}
\usepackage{subcaption}
\usepackage[colorlinks=true,citecolor=darkred,urlcolor=darkred, pdfborder={0 0 0}]{hyperref}
\usepackage[normalem]{ulem}
\definecolor{darkred}{rgb}{0.6,0,0}

\newcommand{\beq}{\begin{equation}}
\newcommand{\eeq}{\end{equation}}
\newcommand{\bea}{\begin{eqnarray}}
\newcommand{\eea}{\end{eqnarray}}
\def\beq#1\eeq{\begin{equation}#1\end{equation}}
\def\bal#1\eal{\begin{align}#1\end{align}}

\newcommand{\beas}{\begin{eqnarray*}}
\newcommand{\eeas}{\end{eqnarray*}}

\def\be{\begin{equation}}
\def\ee{\end{equation}}
\definecolor{linkcolor}{rgb}{0,0,0.5}
\DeclareUnicodeCharacter{2212}{-}
\begin{document}
\bibliographystyle{unsrt} 
\title{
Freeze-in production of sterile neutrino dark matter in a gauged U$(1)^\prime$ model with inverse seesaw  }
\author{Arindam Das }
\email[Email Address: ]{adas@particle.sci.hokudai.ac.jp}
\affiliation{Institute for the advancement of Higher Education, Hokkaido University, Sapporo 060-0817, Japan}
\affiliation{Department of Physics, Hokkaido University, Sapporo 060-0810, Japan}
\author{Srubabati Goswami}
\email[Email Address: ]{sruba@prl.res.in}
\affiliation{Theoretical Physics Division, 
Physical Research Laboratory, Ahmedabad - 380009, India}
\affiliation{Northwestern University, Department of Physics \& Astronomy, 2145 Sheridan Road, Evanston, IL 60208, USA}
\author{Vishnudath K.N.}
\email[Email Address: ]{vishnudath.neelakand@usm.cl}
\affiliation{Universidad T\'{e}cnica Federico Santa Mar\'{i}a, Casilla 110-V, Valpara\'{i}so, Chile}
\author{Tanmay Kumar Poddar}
\email[Email Address: ]{poddar@sa.infn.it}
\affiliation{Istituto Nazionale di Fisica Nucleare-Gruppo Collegato di Salerno-Sezione di Napoli,
Via Giovanni Paolo II, 132 - 84084 Fisciano (SA), Italy}
\begin{abstract}
We consider a general, anomaly free U$(1)^\prime$ extension of the Standard Model (SM) where the neutrino mass is generated at the tree level via the inverse seesaw mechanism. The model contains three right handed neutrinos, three additional singlet fermions, one extra complex scalar and a neutral gauge boson $(Z^\prime)$. Instead of resorting to a specific $U(1)$ extension, we consider a class of models by taking the $U(1)^\prime$ charges of the scalars to be free parameters. Here, we assign one pair of the pseudo-Dirac degenerate sterile neutrinos as Dark Matter (DM) candidates which are produced by the freeze-in mechanism. Considering different mass regimes of the DM, $Z^\prime$ and reheating temperature, we obtain constraints on the $U(1)^\prime$ charges giving the correct relic abundance. We have also obtained constraints on $Z^\prime$ mass and coupling from consideration of relic density as well as high energy collider experiments like ATLAS in case of heavy $Z^\prime$ or in intensity and lifetime frontier experiments like DUNE, FASERs, and ILC beam dump which are looking for light $Z^\prime$. Additionally, in this model, the decay of pseudo-Dirac DM into active neutrinos can explain the 511 keV line observed by the INTEGRAL satellite. 
\end{abstract}
\maketitle
\section{Introduction}
\label{sec1}
The evidence of tiny neutrino mass from the neutrino oscillation experiments \cite{ParticleDataGroup:2020ssz} has heralded the pathway to new physics beyond the Standard Model (SM). Apart from the non-zero neutrino mass, the SM also cannot provide a candidate for the Dark Matter (DM). The non-observation of any signal in the direct detection of DM experiments puts stringent bounds on the DM-nucleon scattering cross section \cite{Aprile:2018dbl,Akerib:2013tjd,Cui:2017nnn}, which severely restrict the models with Weakly Interacting Massive Particles (WIMPs) whose masses are greater than $1$ GeV. There are some scenarios where non-WIMP candidates such as Feebly Interacting Massive Particles (FIMPs) \cite{Hall:2009bx,Blennow:2013jba}, Strongly Interacting Massive Particles (SIMPs) \cite{Hochberg:2014dra}, ultralight gauge bosons \cite{Poddar:2019wvu, Poddar:2020exe,Choi:2020kch}, the axions or Axion Like Particles (ALPs) \cite{Hu:2000ke,Duffy:2009ig,Hui:2016ltb,Poddar:2019zoe,Poddar:2020qft,KumarPoddar:2021ked}, light sterile neutrinos \cite{Boyarsky:2018tvu,Seto:2020udg}, Primordial Black Holes (PBHs) \cite{Carr:2016drx,Lacki:2010zf}, etc. can be potential DM candidates and at the same time can evade the direct detection bounds.  

In this work we extend the SM by a general $U(1)^\prime$ gauge group to explain both the neutrino mass and the DM. We consider the inverse seesaw scenario under $U(1)^\prime$, where the tiny neutrino mass is attributed to a small lepton number violating mass term \cite{Mohapatra:1986bd}. Thus in this model, we have three generations of the SM singlet heavy neutrinos to cancel the gauge and mixed gauge gravity anomalies. In addition to that we have introduced three generations of fermions which are singlets under the SM $\otimes$ U$(1)^\prime$ gauge group. It helps to generate the neutrino mass via the inverse seesaw mechanism. In this scenario, the active light neutrinos are Majorana particles whereas the three pairs of heavy neutrinos are pseudo-Dirac in nature. The model also contains a Beyond Standard Model (BSM) neutral gauge boson whose mass is generated once the $U(1)^\prime$ symmetry is broken spontaneously.

Out of the three pairs of pseudo-Dirac neutrinos, one pair can be considered as the DM candidate whose relic abundance is generated through the freeze-in mechanism \cite{Blennow:2013jba}. In this mechanism, the initial DM abundance is negligibly small and the final DM abundance is obtained by the annihilation of the SM particles. Such a DM candidate is often called as a Feebly Interacting Massive Particle (FIMP), since it couples feebly with the SM particles \cite{Hall:2009bx}. In freeze-in model, the DM yield increases with an increase in the coupling strength in contrast to the thermal freeze-out scenario \cite{Kolb:1990vq,Mohanty:2020pfa}. The freeze-in mechanism is realized in several well motivated frameworks \cite{Shaposhnikov:2006xi,Petraki:2007gq,Abada:2014zra,
Shakya:2015xnx,Barman:2020plp,Barman:2020ifq,Iwamoto:2021fup,Barman:2021tgt}. Depending on whether the DM particles interact with the visible sector via renormalizable or non-renormalizable interactions, one can have infrared (IR) freeze-in~\cite{Hall:2009bx,Cheung:2010gj,Shakya:2015xnx} or ultraviolet (UV) freeze-in \cite{Hall:2009bx,Elahi:2014fsa,
McDonald:2015ljz,Chen:2017kvz} respectively. In the case of UV freeze-in, the DM particles are produced by scattering of the SM particles, mediated by some portal. In an effective field theory prescription, this can be described by a higher dimensional non-renormalizable operator where the cut-off scale is the mass of the mediator. Thus in UV freeze-in, the dominant production of the DM occurs at temperatures greater than the masses of the scattering particles and DM. The production temperature is smaller than the mass of the mediator so that the effective field theory prescription is still valid. On the other hand in IR freeze-in, the DM interacts with the visible sector via a renormalizable operator and the DM production happens at low temperature (temperature down to the  mass of the DM). 

In this paper, we study different scenarios with various ranges for the $Z^\prime$ mass $(M_{Z^\prime})$, DM mass $(m_N)$ and the reheating temperature ($T_R$). In particular, we consider three cases : (case A) $ m_N < T_R < M_{Z^\prime}$, (case B) $M_{Z^\prime} < m_N < T_R$, and (case C) $m_N < M_{Z^\prime} < T_R $. We analyze the constraints on $Z^\prime$ mass and gauge couplings from relic density considerations as well as high energy collider experiment such as ATLAS \cite{ATLAS:2019erb} in the case of heavy $Z^\prime$ and intensity and lifetime frontier experiments  such as FASERs \cite{Ariga:2018uku,Ariga:2019ufm}, ILC beam dump \cite{Asai:2020xnz,Asai:2021ehn}, and DUNE \cite{Dev:2021qjj} in the case of light $Z^\prime$.

Further, we also point out that in the case of MeV scale $m_N$ with TeV scale $Z^\prime$, one can explain the $511$ keV line detected by the INTEGRAL satellite \cite{Knodlseder:2003sv,Jean:2003ci}. Such a solution to the INTEGRAL anomaly was first proposed by authors of \cite{Khalil:2008kp} in the context of a $U(1)_{\mathrm{B-L}}$ model with type I seesaw.


We arrange the paper as follows. In Section.~\ref{sec2} we describe the model. We study the freeze-in production of the 
DM candidate in Section.~\ref{sec3}, followed by the discussions on the different constraints from the experimental searches of a heavy and light $Z^\prime$ in Section.~\ref{sec5}. We discuss the INTEGRAL anomaly in Section.~\ref{inte}.
Finally we conclude the paper in Section.~\ref{sec6}.
\section{The $U(1)^\prime$ inverse seesaw model}
\label{sec2}
\begin{table}[h]
\begin{center}
\small
\begin{tabular}{|c|c|c|c|c|c|c||c|c|c|} \hline
 & $Q_{L_{i}}$ & $u_{R_{i}}$ & $d_{R_{i}}$ & $\ell_{L_{i}}$ & $e_{R_{i}}$ & $\nu_{R_{\alpha}}$ & $H$ & $ \Phi $& $S$ \\ \hline\hline
SU$(3)_{\rm{C}}$ & ${ 3}$ & ${ 3}$ & ${ 3}$ & ${ 1}$ & ${ 1}$ & ${ 1}$ & ${1}$ & ${ 1}$& ${ 1}$ \\ \hline \hline
SU$(2)_{\rm{L}}$ & ${ 2}$ & ${ 1}$ & ${ 1}$ & ${ 2}$ & ${ 1}$ & ${ 1}$ & ${2}$ & ${ 1}$& ${ 1}$ \\ \hline \hline 
U$(1)_{\rm{Y}}$ & $1/6$ & $2/3$ & $-1/3$ & $-1/2$ & $-1$ & $0$ & $1/2$ & $0$&$0$ \\ \hline \hline
$U(1)^\prime$ & $x_{q}$ & $x_u$ & $x_d$& $x_\ell$& $x_e$ & $x_\nu$ & $\frac{x_H}{2}$ & $-x_\Phi$&$0$ \\ \hline
\end{tabular}
\caption{Particle content of the model.}
\label{tab1}
\end{center}
\end{table}
In TABLE \ref{tab1} we demonstrate the particle content of the $U(1)^\prime$ model under consideration. Together with the SM fermions, there are three RHNs ($\nu_R$) and three singlet fermions ($S$). $i$ and $\alpha$ are the generation indices. $H$ is the SM Higgs doublet and $\Phi$ is the complex scalar responsible for the $U(1)^\prime$ symmetry breaking. The $U(1)^\prime$ charges of the SM Higgs and the $U(1)^\prime$ scalar are denoted as $\frac{1}{2}x_H$ and $x_\Phi$, respectively. 
The gauge invariant Yukawa interaction as per the charges shown in TABLE \ref{tab1} is,
\begin{equation}
-\mathcal{L}_{\rm{Yukawa}}=Y_e \overline{\ell_L}H e_R+ Y_\nu\overline{\ell_L} \tilde{H} \nu_R + Y_u \overline{Q_L} \tilde{H} u_R+Y_d \overline{Q_L} Hd_R+ y_{NS} \overline{\nu_R} \Phi S+ \frac{1}{2} \overline{S^c} M_\mu S+\rm{h.c.},
\label{eq:g2}
\end{equation}
where $\tilde{H}=i\sigma_2H^*$ and h.c. denotes the Hermitian conjugate term. The generation indices are suppressed. From the gauge and mixed gauge-gravity anomaly free conditions, one obtains the $U(1)^\prime$ charges of the fermions in terms of $x_H$ and $x_\Phi$ as,
\be x_\ell = -x_\Phi-\frac{x_H}{2}\,\,\,\,,\,\,\,\, x_e = -x_\Phi - x_H  \,\,\,\,,\,\,\,\, x_\nu=-x_\Phi, \nonumber \ee
\be  x_q = \frac{1}{6}(2 x_\Phi + x_H) \,\,\,\,,\,\,\,\, x_u = \frac{1}{3}(2 x_H + x_\Phi) \,\,\,\,, \,\,\,\, x_d= \frac{1}{3}(x_\Phi - x_H).  \ee
Note that the choice $x_\Phi = 1$ and $x_H = 0$ correspond to the U$(1)_{\rm{B-L}}$ model. Further $x_\Phi=1$ and $x_H=-2$ corresponds to the $U(1)_R$ scenario in which only the right handed fermions interact with the $Z^\prime$. Again for $x_\Phi=1$ and $x_H=-1$ there will be no interaction between $e_R$ and $Z^\prime$. Similarly, for $x_\Phi=1$ and $x_H=-0.5$ or $1$, the U$(1)_X$ charges of $u_R$ or $d_R$ become zero respectively forbidding their interactions with the $Z^\prime$. Implications of some of these special cases have been studied at the proposed electron-positron colliders \cite{Das:2021esm} and neutrino experiments \cite{Chakraborty:2021apc}.

After the symmetry breaking, one obtains the neutral fermion mass terms from Eq. \ref{eq:g2} as,
\begin{equation}
-\mathcal{L}_{\rm{mass}}=\overline{\nu_L}M_D\nu_R+\overline{\nu_R}M_R S+\frac{1}{2}\overline{S^c}M_\mu S+\rm{h.c.},
\end{equation}
where, $M_D=Y_\nu\big<H\big>$, and $M_R=y_{NS}\big<\Phi\big>$. One can write the neutral fermion mass matrix as,
\begin{equation}
\mathcal{L}_{\rm{mass}}=\frac{1}{2}\begin{pmatrix}
\overline{\nu^c_L} && \overline{\nu_R}&& \overline{S^c}
\end{pmatrix}\begin{pmatrix}
0 && M^*_D&& 0\\
{M^{\dagger}_D} &&0 && M_R\\
0 &&  M^T_R && M_\mu
\end{pmatrix}\begin{pmatrix}
\nu_L\\
\nu^c_R\\
S
\end{pmatrix},\label{massmatrix}
\end{equation}
where, $M_D$, $M_R$, and $M_\mu$ are $3\times 3$ matrices and they have a mass hierarchy $M_R\gg M_D\gg M_\mu$. In this case, the effective light neutrino mass matrix becomes,
\begin{equation}
M_{\nu}=M^*_D(M^T_R)^{-1}M_\mu M_R^{-1}M^\dagger_D.
\end{equation}
Thus, the smallness of the neutrino mass is attributed to the suppression by both $M_\mu$ and $\frac{M_D}{M_R}$. To obtain light neutrinos of mass $\mathcal{O}(0.1~\textrm{eV})$, one can take $M_\mu$ as $\mathcal{O}(\rm{keV})$ and $\frac{M_D}{M_R}$ as $\mathcal{O}(0.01)$. After diagonalizing the $9\times 9$ neutral fermion mass matrix in Eq.~\ref{massmatrix} following \cite{Grimus:2000vj,Xing:2005kh}, the flavour eigenstates of the light active neutrinos $(\nu_\alpha)$ can be expressed as a linear combination of the mass eigenstates of the light neutrinos $(\nu_i)$ and the heavy neutrinos $(N_j)$ as, 
\begin{equation}
\nu_\alpha = U_{\alpha i}\nu_i+V_{\alpha j}N_j, \label{mixingmatrix}
\end{equation} 
where $U$ is the Pontecorvo-Maki-Nakagawa-Sakata (PMNS) matrix and $V$ is the $3\times 6 $ active-sterile mixing matrix. The heavy neutrino sector consists of three pairs of pseudo-Dirac neutrinos with masses $M_R \pm M_\mu$. It is well known that the minimal inverse seesaw model that can explain the oscillation data requires at least two generations of $\nu_R$ and $S$ respectively \cite{Abada:2014vea}. We use this fact and assume that only two pseudo-Dirac pairs having mass of the $\mathcal{O}(\textrm{TeV)}$ are mainly responsible for the light neutrino masses, making the lightest active neutrino almost massless. The other pair of pseudo-Dirac neutrinos, which we denote as $N_1$ and $N_2$, are the DM candidates. This pair has very small values of Yukawa couplings and hence very small mixing with the active neutrinos as well as the other heavy neutrinos. 

\begin{sidewaystable}
   $$
 \begin{array}{|c|c|c|}

 \hline 
 
 \textrm{Parameters}& \textrm{BP-I} & \textrm{BP-II}  \\

 \hline
 
 Y_\nu & \begin{pmatrix}
(1.066) 10^{-17} & 0.103 - I~0.016 & 0.037 + I~0.039\\
(-3.358  + I~ 0.381) 10^{-18}   &  -0.149 - I~ 0.037  &  0.391 - I~0.028\\
(6.686+I~ 0.309) 10^{-18} &  -0.236 - I ~ 0.012 & 0.141 - I~0.047
\end{pmatrix}   &   \begin{pmatrix}
(1.221) \times 10^{-20}   & 0.086 + I~0.003 & -0.040 + I~0.026\\
(-4.812 + I~0.356) 10^{-21}  &  0.071 - I~ 0.028  &  0.251 + I~0.009 \\
(6.815 + I~0.382) 10^{-21}  &  -0.101 - I~ 0.027  &  0.250 - I~0.013 
\end{pmatrix} \\
 
 \hline
 
 Tr[Y_\nu^\dag Y_\nu]  & 0.269  & 0.152 \\
 
 \hline
 
 M_R (\textrm{GeV})& \begin{pmatrix}
0.001&0&0\\
0 &8625.978&0\\
0&0&6463.302
\end{pmatrix}&\begin{pmatrix}
4747.656&0&0\\
0 &7005.899&0\\
0&0&4095.693
\end{pmatrix} \\
 \hline
  M_\mu  ~(\textrm{GeV}) &  \begin{pmatrix}
1.945 \times 10^{-8} & 0 & 0\\
0  & 8.660 \times 10^{-7}  &  0\\
0 &  0 & 2.302 \times 10^{-7}
\end{pmatrix}  &  \begin{pmatrix}
3.401\times 10^{-7} & 0 & 0\\
0  & 6.850 \times 10^{-7}  &  0\\
0 &  0 & 2.159\times 10^{-7}
\end{pmatrix} \\
 
 \hline
 
 \delta_{CP}  & 3.438 & 3.438 \\

\hline

\alpha, \beta  &  0,0   &  0,0 \\

\hline

m_{\textrm{light}} ~(\textrm{eV}) &  10^{-22},  8.476\times 10^{-3},~5.056\times 10^{-2}  & 10^{-40},  8.802\times 10^{-3}, ~4.955\times 10^{-2}  \\
 
 \hline
 
 \Sigma_\alpha |V_{\alpha 1}|^2  & 2.570 \times 10^{-24}  &  1.470\times 10^{-43} \\
 
 \hline
 
 \end{array}
 $$
\caption{Benchmark neutrino parameters that satisfy all experimental bounds discussed in the text for the NH case.}
\label{tab2}
\end{sidewaystable}
 
In TABLE \ref{tab2}, we have given two sample Benchmark Points (BP) for which constraints obtained from oscillation experiments \cite{Esteban:2020cvm,nufit}, 
non unitarity of the PMNS matrix \cite{Antusch:2016brq}, lepton flavour violating processes \cite{Bertl:2006up,Coy:2018bxr}, and X-ray observations \cite{Perez:2016tcq,RiemerSorensen:2009jp,Horiuchi:2013noa,Loewenstein:2008yi,Watson:2006qb,Yuksel:2007xh,Mirabal:2010an} are satisfied. 
For simplicity, we have taken both $M_R$ and $M_\mu$ to be diagonal and real. Taking very small mixing of the DM candidates with the active neutrinos forces the lightest active neutrino to take extremely small values as shown in TABLE \ref{tab2}. 
\section{Sterile neutrino dark matter}
\label{sec3}
We have a pair of almost degenerate pseudo-Dirac neutrinos which act as a DM candidate. The corresponding mass terms are given as,
\begin{equation} \label{mat1}
-\mathcal{L}_{DM mass} = \frac{1}{2}
\begin{pmatrix}
 \bar{\nu_R}_1 && \bar{S^c}_1
\end{pmatrix}\begin{pmatrix}
 0 && {M_R}_{11}\\
   {M_R}_{11} && {M_\mu}_{11}
\end{pmatrix}\begin{pmatrix}
{\nu^c_R}_1\\
S_1
\end{pmatrix}. 
\end{equation}
The mass matrix can be diagonalized by $U=\mathcal{O}\rho$, where $\mathcal{O}$ is an orthogonal $2\times 2$ matrix given by,
\begin{equation}
\mathcal{O}=\begin{pmatrix}
\cos\theta & \sin\theta\\
-\sin\theta &\cos\theta 
\end{pmatrix},
\label{mat3}
\end{equation}
and $\rho$ is a diagonal phase matrix given by $\rho=\mathrm{diag}(\rho_1,\rho_2)$ with $\rho_k^2=\pm1$, where $k=1,2$. The matrix $\mathcal{O}$ diagonalizes the mass matrix in Eq. \ref{mat1} such that $\mathcal{O}^T M\mathcal{O}=diag(m^\prime_1,m^\prime_2)$, where $m^\prime_{1,2}$ are the eigenvalues of the mass matrix given as,
\begin{equation}
m^\prime_{2,1}=\frac{1}{2}[{m_\mu}_{11}\pm \sqrt{m^2_{\mu 11}+4m^2_{R11}}].
\label{mat2}
\end{equation} 
The phase matrix $\rho$ ensures that the mass eigenvalues are positive and real. Thus, $U^T MU=\rho^T\mathcal{O}^TM\mathcal{O}\rho=\mathrm{diag}(\rho^2_1m^\prime_1,\rho^2_2m^\prime_2)$, and $\tilde{m}_k=\rho^2_km^\prime_k$. Hence, for $\rho^2_1=-1$ and $\rho_2^2=1$, the two mass eigenvalues become,
\begin{equation}
\tilde{m}_{2,1}=\frac{1}{2}\sqrt{m^2_{\mu 11}+4m^2_{R11}}\pm\frac{m_{\mu11}}{2}.
\label{mat3}
\end{equation}
In this scenario, the mass square difference is very small and is given as $\Delta \tilde{m}^2=\tilde{m}^2_2-\tilde{m}^2_1=m_\mu\sqrt{m^2_\mu+4m_R^2}$ and the mixing angle is defined as $\tan2\theta=2m_{R11}/m_{\mu11}$. 
Thus, one can express the flavor eigenstates $(\nu_R^c, S_1)$ in terms of the mass eigenstates $(N_1, N_2)$ as,
\begin{equation}
\nu_{R1}=i(N_1\cos\theta-N_2\sin\theta)~~~ \&~~~ S_1=N_1\sin\theta+N_2\cos\theta.
\label{mat5}
\end{equation}
These pair of massive pseudo-Dirac neutrinos $(N_1, N_2)$ are the DM candidates in our model. We consider the non thermal production of sterile neutrino DM satisfying the relic abundance by freeze-in process through $Z^\prime$ 
portal. In this scenario, the initial number density of the DM is very small and the abundance is built up by the DM interactions with the SM thermal bath. However, the interaction rate is assumed to be very small so that the DM particles do not attain thermal equilibrium with the bath. Therefore, the DM annihilation to the SM particles can be ignored. In the context of the $U(1)^\prime$ scenario, the DM particles are produced by the annihilation of the SM particles mediated by the $Z^\prime$ gauge boson. The interaction of the $Z^\prime$ with the DM is given as,
\begin{equation}
\begin{split}
-\mathcal{L}=g^\prime x_\nu {Z^\prime}^\mu\bar{\nu}_{R1}\gamma_\mu \nu_{R1}=g^\prime x_\nu {Z^\prime}^\mu(\cos^2\theta\bar{N_1}\gamma_\mu N_1+\sin^2\theta\bar{N_2}\gamma_\mu N_2\\
-\sin\theta\cos\theta\bar{N_1}\gamma_\mu N_2
-\sin\theta\cos\theta\bar{N_2}\gamma_\mu N_1),
\end{split}
\label{csk}
\end{equation}
where $x_\nu=-x_\Phi$. Similarly, the $Z^\prime$ interactions with the SM fermions are written as, 
\begin{equation}
-\mathcal{L}=g^\prime Q_L {Z^\prime}^\mu \bar{f_L}\gamma_\mu f_L+g^\prime Q_R {Z^\prime}^\mu\bar{f_R}\gamma_\mu f_R,
\end{equation}
with $Q_L$ and $Q_R$ being the $U(1)^\prime$ charges of the left and right chiral fermions. 

Since we have a pair of pseudo-Dirac neutrinos $(N_1, N_2)$ as DM candidate, we can write the DM yields for the freeze-in scattering process as,
\begin{equation}
Y^\prime_{N_1}\simeq\frac{\mathcal{S}}{Hx^2_1}[\langle \sigma v(ff\rightarrow N_2N_1)\rangle Y_{\rm{EQ}}^{N_2}Y_{\rm{EQ}}^{N_1}+\langle \sigma v(ff\rightarrow N_1N_1)\rangle {Y_{\rm{EQ}}^{N_1}}^2],
\end{equation}
and 
\begin{equation}
Y^\prime_{N_2}\simeq\frac{\mathcal{S}}{Hx^2_2}[\langle \sigma v(ff\rightarrow N_1N_2)\rangle Y_{\rm{EQ}}^{N_1}Y_{\rm{EQ}}^{N_2}+\langle \sigma v(ff\rightarrow N_2N_2)\rangle {Y_{\rm{EQ}}^{N_2}}^2],
\end{equation}
where $\mathcal{S}$ and $H$ denote the entropy density and Hubble parameter respectively and $\langle \sigma v\rangle$ denotes the thermal average product of the cross section and the relative velocity of the initial state particles. The prime denotes the first order derivative with respect to $x_{1,2}=\frac{m_{N_{1,2}}}{{T}}$, where $T$ is the temperature of the universe at the radiation dominated epoch. We consider that the initial DM abundance to be zero. The DM abundance in the universe gradually grows from the scattering of SM particles. As the expansion rate of the universe becomes greater than the interaction rate of $ff\rightarrow \rm{DM}~\rm{DM}$, the DM yield freezes-in to the present DM relic density of the universe.

In this work, we consider the maximal mixing scenario which corresponds to the case $m_R\gg m_\mu$ and $\theta=45^{\circ}$. Therefore, Eq. \ref{mat3} becomes,
\begin{equation}
\tilde{m}_{2,1}\simeq m_{R11}\pm\frac{m_{\mu11}}{2}\simeq m_{R11}= m_N,
\end{equation}
so that we can avoid the possible decay of one sterile neutrino into another. Hence, both $N_1$ and $N_2$ contribute to the DM relic density. Therefore, the total DM yield becomes,
\begin{equation}
\begin{split}
Y^\prime_{\rm{tot}}=Y^\prime_{N_1}+Y^\prime_{N_2}\simeq \frac{\mathcal{S}}{Hx^2}[\langle \sigma v(ff\rightarrow N_1N_1)\rangle {Y_{\rm{EQ}}^{N_1}}^2+\langle \sigma v(ff\rightarrow N_2N_2)\rangle {Y_{\rm{EQ}}^{N_2}}^2\\
+2\langle \sigma v(ff\rightarrow N_1N_2)\rangle Y_{\rm{EQ}}^{N_1}Y_{\rm{EQ}}^{N_2}]=\frac{\mathcal{S}}{Hx^2}\langle \sigma v(ff\rightarrow NN)\rangle {Y_{\rm{EQ}}^{\rm{tot}}}^2,
\end{split}
\label{cd1}
\end{equation}
since $\langle \sigma v(ff\rightarrow N_1N_1)\rangle\simeq\langle \sigma v(ff\rightarrow N_2N_2)\rangle\simeq\langle \sigma v(ff\rightarrow N_1N_2)\rangle=\langle \sigma v(ff\rightarrow NN)\rangle$, and $x_1\simeq x_2\simeq x=\frac{m_N}{T}$ which follows from $m_{N_1}\simeq m_{N_2}=m_N$. Thus, $Y^{\rm{tot}}_{\rm{EQ}}=Y^{N_1}_{\rm{EQ}}+Y^{N_2}_{\rm{EQ}}\simeq 2Y^{N}_{EQ}$. The initial condition for the production of DM by freeze-in mechanism is $Y(x_{i})=0$ for $x_i\ll1$ and the coupling between the DM and the SM particles is very small.

Note that in our case, the elastic scattering processes $ff\rightarrow N_{1(2)}N_{1(2)}$ contribute to the DM yield by the same amount as the inelastic scattering $ff\rightarrow N_1N_2$. This is due to the fact that in the flavour basis, only $\nu_{R1}$ has a gauge interaction whereas $S_1$ is gauge singlet (Eq. \ref{csk}) as well as $\theta=45^\circ$. The elastic scattering contributions will be cancelled in a scenario where the $U(1)^\prime$ charges of $\nu_{R1}$ and $S_1$ are equal and opposite \cite{CarrilloGonzalez:2021lxm}. In this paper, our emphasis is on exploring the phenomenology of DM through the $Z^\prime$ portal within the framework of the $U(1)^\prime$ model. Hence, we have excluded the contribution due to the scalar-mediated, and scalar-decay diagrams in the DM production process. This corresponds to taking the quartic coupling between the SM Higgs and the $U(1)^\prime$ scalar to be very small and taking large mass for the scalar. \cite{Shaposhnikov:2006xi,Kusenko:2006rh,Petraki:2007gq,Matsui:2015maa}. Also, if the mass of the $Z^\prime$ is greater than the DM mass, the decay of $Z^\prime$ to DM particles will be important for DM production. However, here we assume that the $Z^\prime$ is not thermalized with the thermal bath and the scattering processes are the dominant channels for the DM population. If the mass of the $Z^\prime$ is greater than the DM mass, and if $Z^\prime$ is in thermal equilibrium, then the decay of $Z^\prime$ to DM particles will be important for DM relic density contribution \cite{McDonald:2001vt,Covi:2002vw,Hall:2009bx,Yaguna:2011qn,Li:2023ewv,Allahverdi:2024gdv}

In the following, we discuss the non-thermal production of the sterile neutrino DM via the freeze-in mechanism for different mass regimes of $Z^\prime$ through scattering processes. We calculate the DM relic density in each case and constrain the model parameters by requiring the sterile neutrino DM candidate to account for the entire relic abundance of the universe.
Particularly, we consider three cases : 
\begin{itemize}
\item \textbf{Case A:} $ m_N < T_R < M_{Z^\prime}$ with $M_{Z^\prime} \sim \mathcal O$(TeV), $10~\mathrm{keV} \leq m_N \leq 1 ~\mathrm{MeV}$ and   $ 5~\mathrm{MeV} \leq T_R \leq 100 ~\mathrm{MeV} $   
\item\textbf{Case B:} $ M_{Z^\prime} < m_N < T_R $ with  $10 ~\mathrm{MeV} \leq M_{Z^\prime} \leq 5~\mathrm{GeV}$, $m_N \sim \mathcal{O}$(TeV) and $T_R \ge \mathcal{O}$(TeV)  and 
\item\textbf{Case C:} $m_N <  M_{Z^\prime} < T_R $ with 10 MeV $\leq M_{Z^\prime} \leq$ 5 GeV,  10 keV $\leq m_N \leq $ 100 keV and $T_R \ge \mathcal O$(TeV). 
\end{itemize}

\subsection{Case A: $M_Z^\prime> T_R> m_N$}
\label{subsec4}
In the heavy $Z^\prime$ case with $M_{Z^\prime}\sim \mathcal{O}(\mathrm{TeV})$, the pair of pseudo-Dirac DM ($\mathrm{keV} \leq m_N \leq \mathrm{MeV}$) is produced by scattering of the SM particles mediated by $Z^\prime$. Since, we are considering low reheating temperature ($5~\mathrm{MeV} \leq T_R \leq 100~\mathrm{MeV}$) the only relevant processes for the production of DM are through the scattering of electron-positron $(e^+e^-)$ and neutrino-antineutrino $(\nu\bar{\nu})$ pairs.

We calculate the final abundance of the freeze-in DM yield from the Boltzmann equation as,
\begin{equation}
Y_\infty=\int^{T_R} \frac{<\sigma v(f\bar{f}\rightarrow NN)>n^2_\mathrm{eq}}{\mathcal{S}HT}dT,
\label{eq:u9}
\end{equation}
where $n_\mathrm{eq}$ denotes the equilibrium number density of DM. The Hubble parameter in the radiation dominated epoch is $H(T)=1.66\sqrt{g_*}\frac{T^2}{M_{\rm{pl}}}$ and the entropy density is given as $\mathcal{S}(T)=\frac{2\pi^2}{45}g_{*}T^3$. Here, $M_{pl}$ is the reduced Planck mass and $g_*=10.75$ is the total number of effective massless degrees of freedom. 

In the massless limit of the initial state particles, we calculate the spin averaged amplitude square for the process $e^+e^-\rightarrow NN$ in the $s$ channel mediated by the $Z^\prime$ as,
\begin{equation}
\begin{split}
\int \frac{d\cos\theta}{2}|\bar{\mathcal{M}}_{e^+e^-}(s,\cos\theta)|^2=\frac{{g^\prime}^4 {x^2_\Phi/4}}{(s-M^2_{Z^\prime})^2+M^2_{Z^\prime}\Gamma^2_{Z^\prime}}\times\frac{2}{3}s(s-4m^2_N)\Big[\Big(\frac{3}{4}x_H+x_\Phi\Big)^2\\
+\Big(\frac{x_H}{4}\Big)^2\Big].
\end{split}
\label{eq:u2}
\end{equation}
Similarly, for $\nu\bar{\nu}\rightarrow NN$, we obtain,
\begin{equation}
\begin{split}
\int \frac{d\cos\theta}{2}|\bar{\mathcal{M}}_{\nu\bar{\nu}}(s,\cos\theta)|^2=\frac{{g^\prime}^4 x^2_\Phi/4}{(s-M^2_{Z^\prime})^2+M^2_{Z^\prime}\Gamma^2_{Z^\prime}}\times\frac{1}{3}s(s-4m^2_N)\Big(x_\Phi+\frac{x_H}{2}\Big)^2,
\end{split}
\label{eq:u3}
\end{equation}
where $\sqrt{s}$ denotes the centre of mass energy for the scattering and $\Gamma_{Z^\prime}$ denotes the decay width of $Z^\prime$. 

Thus, Eq. \ref{eq:u9} becomes,
\begin{equation}
\begin{split}
Y_\infty\simeq 3.5\times 10^{-7}\times\Big(\frac{10}{g_*}\Big)^\frac{3}{2}\Big(\frac{T_R}{5~\rm{MeV}}\Big)^3\Big(\frac{9.7~\rm{TeV}}{M_{Z^\prime}/g^\prime}\Big)^4\times \frac{x^2_\Phi}{6}\Big[2\Big\{\Big(\frac{3}{4}x_H+x_\Phi\Big)^2+\Big(\frac{x_H}{4}\Big)^2\Big\}\\
+\Big(x_\Phi+\frac{x_H}{2}\Big)^2\Big].
\end{split}
\label{eq:u10}
\end{equation}
The density parameter for the DM is given by $\Omega_N=\frac{\rho}{\rho_c}$, where $\rho_c=\frac{3H_0^2}{8\pi G}=1.05\times 10^{-5}h^2~\mathrm{GeV}/\mathrm{cm^3}$, $h$ is the reduced Hubble constant, $\mathcal{S}_0=2889.2~\mathrm{cm^{-3}}$ and $\rho=m_N \mathcal{S}_0Y_\infty$. Hence, the abundance of the DM becomes,
\begin{equation}
\begin{split}
\Omega_Nh^2\simeq0.12\times \Big(\frac{m_N}{1~\rm{MeV}}\Big)\Big(\frac{10}{g_*}\Big)^\frac{3}{2}\Big(\frac{T_R}{5~\rm{MeV}}\Big)^3\Big(\frac{9.7~\rm{TeV}}{M_{Z^\prime}/g^\prime}\Big)^4\frac{x^2_\Phi}{6}\Big[2\Big\{\Big(\frac{3}{4}x_H+x_\Phi\Big)^2+\Big(\frac{x_H}{4}\Big)^2\Big\}\\
+\Big(x_\Phi+\frac{x_H}{2}\Big)^2\Big].
\end{split}
\label{Eq:u11}
\end{equation}
The relic density of DM increases with increase in the mass of the DM and the reheating temperature and with decrease in the $U(1)^\prime$ gauge boson mass. The relic density of the DM also increases with increase in the $U(1)^\prime$ charges of the two scalars in the model.
\begin{figure}
\centering
\begin{subfigure}{0.51\textwidth}
    \includegraphics[width=\textwidth]{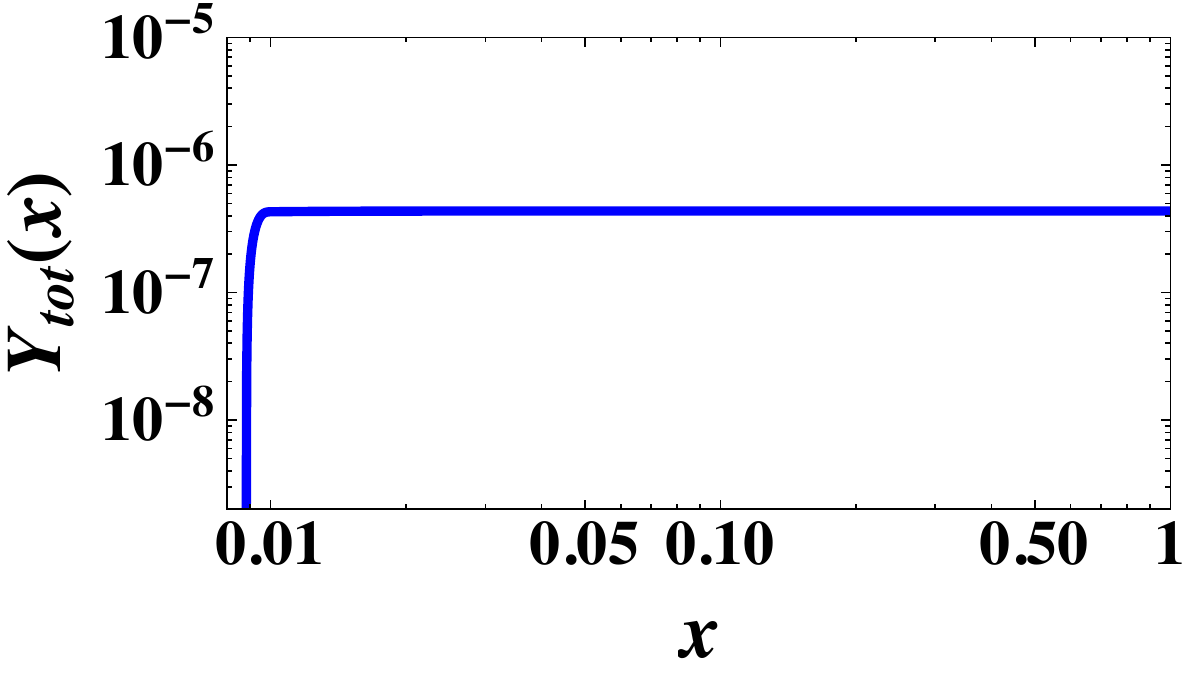}
    \caption{$Y(x)$ vs. $x$}
    \label{fig:first}
\end{subfigure}
\begin{subfigure}{0.48\textwidth}
    \includegraphics[width=\textwidth]{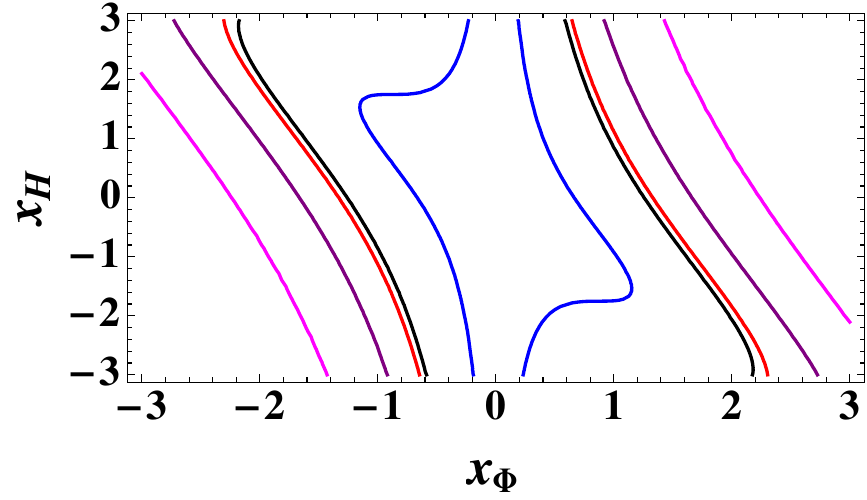}
    \caption{$x_H$ vs. $x_\Phi$}
    \label{fig:second}
\end{subfigure}
\caption{(a) DM yield $Y(x)$ vs. $x=\frac{m_N}{T}$ for the case A $(M_Z^\prime> T_R> m_N)$. (b) Contour plots of $x_H$ vs. $x_\Phi$ that satisfy the correct relic density $\Omega_N h^2=0.12$ for case A.}

\label{Fig:mkk1}
\end{figure}

In Fig.~\ref{fig:first} we have shown the variation of the DM yield $Y(x)$ with $x$ by numerically solving the Boltzmann equation. In plotting this, we have fixed $m_N=1~\rm{MeV}$, $g^\prime=0.15$, $M_{Z^\prime}=10~\rm{TeV}$, $T_R=100~\rm{MeV}$, $x_H=-1.2$\footnote{Note that, for this choice of $x_H$, the $Z^\prime$ branching ratio into a pair of RHNs is dominant over that into the dilepton unlike the $U(1)_\mathrm{B-L}$ scenario \cite{Das:2017flq}.}, $x_\Phi=1$, and the resultant $Y_\infty=4.36\times 10^{-7}$ gives the correct relic density. The DM freezes in for $x\simeq 0.0088$.

In Fig.~\ref{fig:second} we obtain the contour plots in the $x_H-x_\Phi$ plane. Along the contour lines, the DM satisfies the correct relic density of the universe. We obtain these contour lines by analytically solving the Boltzmann equation for heavy $Z^\prime$ case. The blue, red, purple, black, and magenta lines correspond to $(T_R, M_{Z^\prime}, g^\prime, m_N)=$ (100 MeV, 5 MeV, 0.1, 1 MeV), (100 MeV, 10 TeV, 0.1, 1 MeV), (100 MeV, 10 TeV, 0.08, 1 MeV), (50 MeV, 10 TeV, 0.1, 10 MeV), (50 MeV, 10 Tev, 0.1, 1 MeV) respectively. Note that for any given value of $x_H$, there exist two values of $x_\Phi$ that give the same relic density. For instance, when $x_H=0$, $x_\phi=\pm n$ give the same relic density for any real number $n$.

The value of the $U(1)^\prime$ gauge coupling that satisfies the correct relic density requirement can be related to that in $U(1)_{\mathrm{B-L}}$ model as,
\begin{equation}
g^\prime=1.32x^{-\frac{1}{2}}_\Phi g_{\mathrm{B-L}}\Big[2\Big\{\Big(\frac{3}{4}x_H+x_\Phi\Big)^2+\Big(\frac{x_H}{4}\Big)^4\Big\}+\Big(x_\Phi+\frac{x_H}{2}\Big)^2\Big]^{-\frac{1}{4}}.
\end{equation}
For $T_R=100~\mathrm{MeV}$, $m_N=1~\mathrm{MeV}$, and $M_{Z^\prime}=10~\mathrm{TeV}$, the $U(1)_{\mathrm{B-L}}$ coupling becomes $g_{\mathrm{B-L}}=0.133$. In the limit $x_H\gg x_\Phi$ and $x_\Phi=1$, we obtain the expression for $g^\prime$ for the above parameters as $g^\prime=\frac{0.16}{\sqrt{x_H}}$. Thus, in this case the $U(1)^\prime$ coupling largely depends on the choice of $x_H$ and the result largely deviates from the $U(1)_\mathrm{B-L}$ case.

Note that, even though we have restricted ourselves to low $T_R$ in this particular scenario, it is also possible to have values of $T_R$ as high as $\mathcal{O}(\rm{TeV})$, the only requirement being that $M_{Z^\prime}>T_R$ so that the effective field theory description will work. For large $T_R (\gtrsim 300~\mathrm{GeV})$, all the SM particles are involved in the scattering process and one needs to choose a small value of $g^\prime$ to prevent overabundance of the DM relic density.

Though the large $T_R$ limit is conducive to baryogenesis via leptogenesis, it is also possible to have baryogenesis in the low $T_R$ limit if we supersymmetrize the model \cite{Khalil:2007dr} and use the Affleck-Dine Mechanism \cite{Affleck:1984fy,Stewart:1996ai,Campbell:1998yi,Dolgov:2002vf}.
\subsection{Case B: $M_Z^\prime< m_N < T_R$}
\label{subsec5}
Now we consider the light $Z^\prime$ scenario, particularly $M_{Z^\prime}$ in the range $10~\mathrm{MeV}\leq M_{Z^\prime}\leq 5~\mathrm{GeV}$ and $T_R> m_N> M_{Z^\prime}$. Here, because of the kinematic constraints, the production of the DM from the thermal plasma almost stops when the temperature becomes smaller than $2m_N$. In this limit, we can neglect $M_{Z^\prime}$ and $\Gamma_{Z^\prime}$ in the denominator of the amplitude squared term for $ff\rightarrow NN$ processes and the cross section in the context of high $T_R$ becomes,
\begin{equation}
\sigma(s)=\frac{1}{48\pi}\frac{g^{\prime4}s\Big(1-\frac{4m^2_N}{s}\Big)^\frac{3}{2}}{s^2}\frac{x^2_\Phi}{4}(10 x^2_H+13 x^2_\Phi+16 x_H x_\Phi).
\label{Eq:alpha2}
\end{equation}
For high temperature $(T\gtrsim 300~ \rm{GeV})$, all the SM particles are relativistic, and the total relativistic degrees of freedom becomes $g_*=106.75$. Hence for $x\lesssim 1$ we can write the Boltzmann equation from Eq. \ref{cd1} as,
\begin{equation}
\frac{dY_\mathrm{tot}}{dx}\simeq \frac{2.8}{g_*^\frac{3}{2}}M_{pl}m_N\frac{4<\sigma v>}{x^2}.
\label{Eq:alpha3}
\end{equation}
The thermal averaged DM annihilation cross section can be written as, 
\begin{equation}
<\sigma v>=\frac{1}{(n_{eq})^2}g^2_N\frac{m_N}{64\pi^2x}\int^\infty_{4m^2_N}ds\times 2(s-4m^2_N)\sigma(s)\sqrt{s}K_1{\Big(\frac{\sqrt{s}}{T}\Big)},
\label{Eq:beta1}
\end{equation} 
where $n_{\mathrm{eq}}=\frac{g_N m^3_N}{2\pi^2 x}K_2(x)$ is the equilibrium number density of DM. Here, $g_N=2$ is the degree of freedom of sterile neutrino DM and $K_2(x)$ is the modified Bessel function of the second kind. If the DM production is mediated by light $Z^\prime$ and $m_N > M_Z^\prime$, then for $x\lesssim 1$,  Eq.~\ref{Eq:beta1} becomes,
\begin{equation}
<\sigma v>\simeq \frac{g^{\prime4}}{384\pi}\frac{x^2}{m^2_N}\frac{x^2_\Phi}{4}(10x^2_H +13x^2_\Phi+16 x_H x_\Phi).
\label{Eq:alpha4}
\end{equation}
Since, the production of the DM particle from thermal plasma stops at $x\simeq1$ (i.e; $T\simeq m_N$), the DM yield is approximated as $Y_{\infty}=Y_\mathrm{tot}(x\simeq 1)$. Substituting Eq.~\ref{Eq:alpha4} in the Boltzmann equation (Eq.~\ref{Eq:alpha3}) and integrating from $x_i\ll1$ to $x=1$, we obtain the analytical solution of the DM yield as,
\begin{equation}
Y_\infty\simeq 2.32\times 10^{-3}\frac{g^{\prime4}}{g_*^\frac{3}{2}}\frac{M_{pl}}{m_N}x^2_\Phi(10x^2_H+13x^2_\Phi+16x_H x_\Phi)\hspace{0.5cm} \text{at}\hspace{0.3cm} T\simeq m_N.
\end{equation}
Hence, the comoving yield of the sterile neutrino DM is inversely proportional to its mass. We can write the relic density of $m_N$ as
\begin{equation}
\Omega_N h^2=\frac{m_N Y_\infty \mathcal{S}_0}{\frac{\rho_c}{h^2}}=0.12\times \Big(\frac{106.75}{g_*}\Big)^\frac{3}{2} \Big(\frac{g^\prime}{3.04\times 10^{-6}}\Big)^4 x^2_\Phi(10x^2_H+13x^2_\Phi+16x_H x_\Phi).
\label{Eq:a8}
\end{equation} 
The analytical result shows that relic density of DM in the light $Z^\prime$ case is independent of the masses of $Z^\prime$, DM, and also on $T_R$. It only depends on the $U(1)^\prime$ gauge coupling and on the scalar charges. The relic density rises with increasing the values of $g^\prime, x_H$ and $x_\Phi$. As before, substituting $x_H=0$ and $x_\Phi=1$ gives us the results for the $U(1)_{\mathrm{B-L}}$ scenario. This case is similar to the IR freeze-in scenario. Here also, corresponding to each value of $x_H$, there exists two values of $x_\Phi$ that give the same value for the relic density. As in Case A, for $x_H=0$, $x_\Phi=\pm n$ give the same value for the DM relic density.
\begin{figure}
\centering
\begin{subfigure}{0.50\textwidth}
    \includegraphics[width=\textwidth]{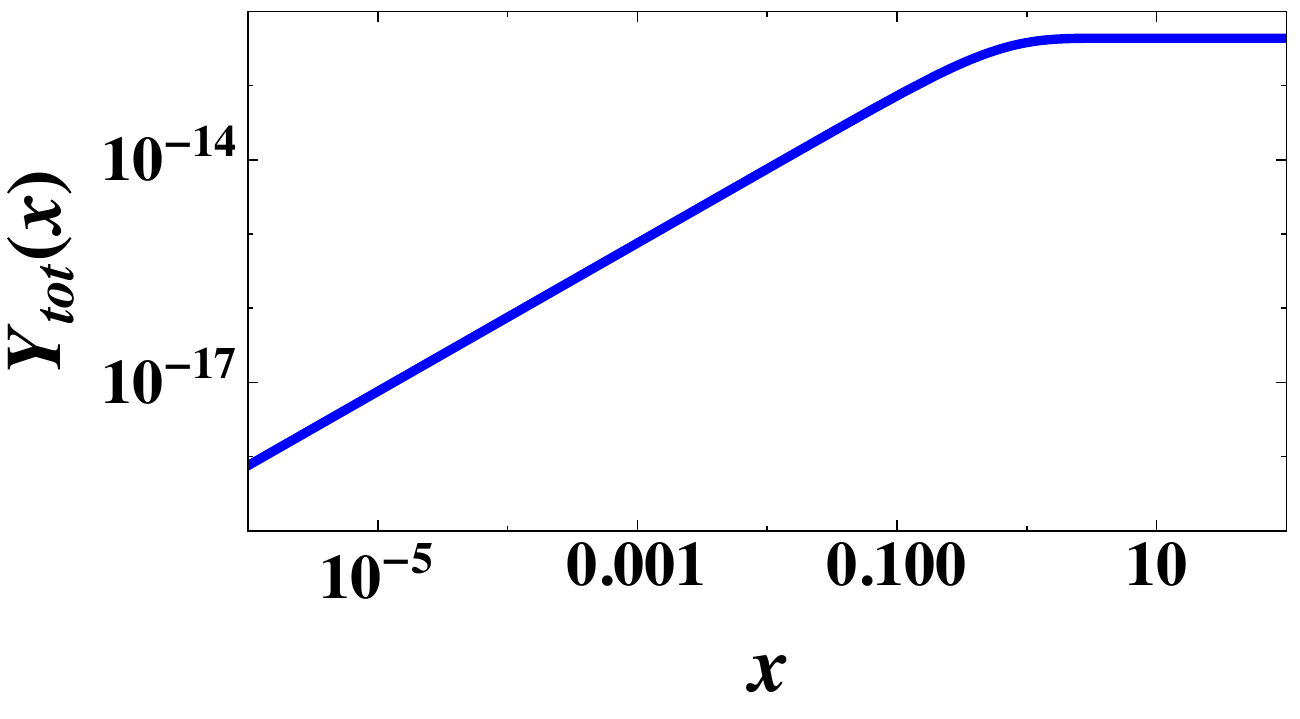}
    \caption{$Y(x)$ vs. $x$}
    \label{fig:lightfirst}
\end{subfigure}
\begin{subfigure}{0.48\textwidth}
    \includegraphics[width=\textwidth]{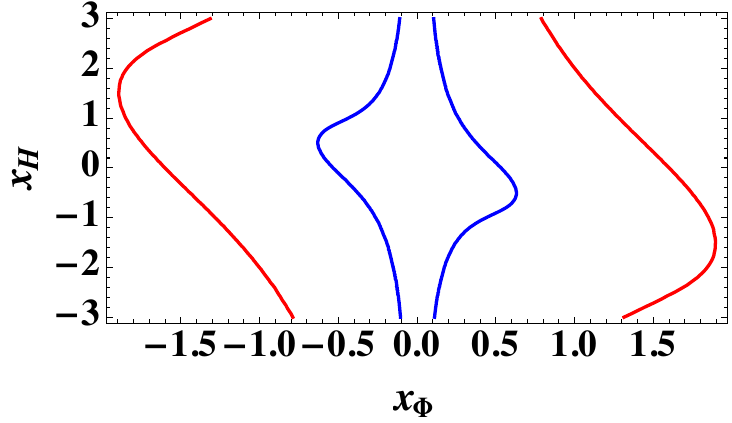}
    \caption{ $x_H$ vs. $x_\Phi$ }
    \label{fig:lightsecond}
\end{subfigure}
\caption{(a) DM yield $Y(x)$ vs. $x=\frac{m_N}{T}$ for case B $(M_{Z^\prime}<m_N<T_R)$. (b) Contour plots of $x_H$ vs. $x_\Phi$ that satisfy the correct relic density $\Omega_N h^2=0.12$ for case B.}
\label{Fig:mkk2}
\end{figure}

In FIG.~\ref{fig:lightfirst}, we numerically solve the Boltzmann equation with the initial condition $Y(x_i)=0$ for $x_i=10^{-9}$ and obtain the DM yield as a function of $x$ for Case B. Here, we have chosen $m_N=1~\mathrm{TeV}, M_{Z^\prime}=0.1~\mathrm{GeV}, x_H=-1.2, x_\Phi=1$ and $g^\prime=3.65\times 10^{-6}$. The freeze-in occurs at $T\sim m_N$ and the resultant $Y_\infty=4.36\times 10^{-13}$ reproduces the observed relic density of DM in the universe. 

In FIG.~\ref{fig:lightsecond}, we have shown the contour lines in the $x_H-x_\Phi$ plane which reproduce the correct relic density. The red and the blue lines correspond to $g^\prime=10^{-6}$ and $g^\prime=3\times 10^{-6}$ respectively. One can take large values of $g^\prime$ to satisfy the DM relic density by considering smaller values of $x_H$ and $x_\Phi$.

For the $U(1)_{\mathrm{B-L}}$ model, the correct relic density of DM is obtained for $g_{\mathrm{B-L}}=1.6\times 10^{-6}$. For the points that give $\Omega_Nh^2=0.12$, the $U(1)^\prime$ and $U(1)_{\mathrm{B-L}}$ gauge couplings are related with each other as,  
\begin{equation}
g^\prime = 1.89 \times {g_{B-L}} x_\Phi^{-\frac{1}{2}}(10x^2_H+13x^2_\Phi+16x_H x_\Phi)^{-\frac{1}{4}}.
\label{Eq:aw8}
\end{equation} 
In the limit $x_H\gg x_\Phi$, and $x_\Phi=1$, Eq. \ref{Eq:a8} becomes $g^\prime=\frac{1.7\times 10^{-6}}{\sqrt{x_H}}$. Therefore, the $U(1)^\prime$ coupling can be made sizeable with the suitable choice of $x_H$.

\subsection{Case C: $m_N< M_{Z^\prime}< T_R$}
\label{int}
Now we consider the case, where the DM is lighter than $Z^\prime$ in the high $T_R$ limit. In this case $m_N< M_{Z^\prime}< T_R$, the DM can be produced through $Z^\prime$ resonance. The resonance occurs at $s=M^2_{Z^\prime}$ for $\sqrt{s}\geq 2m_N$. Using the Breit-Wigner formula for narrow width approximation, the $Z^\prime$ propagator can be written as,
\begin{equation}
\frac{1}{(s-M^2_{Z^\prime})^2+M^2_{Z^\prime}\Gamma^2_{Z^\prime}}=\frac{\pi}{M_{Z^\prime}\Gamma_{Z^\prime}}\delta(s-M^2_{Z^\prime}).
\label{Eq:q2}
\end{equation}
Therefore, the cross section for the resonance production of DM for high $T_R$ becomes,
\begin{equation}
\sigma(s)=\frac{g^{\prime 4}}{48\pi}\pi\frac{s\Big(1-\frac{4m_N^2}{s}\Big)^\frac{3}{2}}{M_Z^\prime\Gamma_{Z^\prime}}\delta(s-M^2_{Z^\prime})\frac{x^2_\Phi}{4}(10x^2_H +13x^2_\Phi+16x_H x_\Phi).
\label{Eq:beta2}
\end{equation}
If the mass of the $Z^\prime$ is such that it cannot decay to top quark ($M_{Z^\prime} < 2 m_t \sim 346$ GeV) then the total decay width of $Z^\prime$ becomes,
\begin{equation}
\Gamma_{Z^\prime}=\frac{1}{24\pi}g^{\prime2}M_Z^\prime[\frac{1}{12}(103x^2_H+148x^2_\Phi+172x_H x_\Phi)+x^2_\Phi].
\label{Eq:a15}
\end{equation}
Using Eqs.~\ref{Eq:beta1}, \ref{Eq:q2}, \ref{Eq:beta2}, and \ref{Eq:a15}, we can write the thermal average cross section as,
\begin{equation}
<\sigma v>\simeq \frac{\pi g^{\prime 2}x^5M^3_{Z^\prime}}{64m^5_N}K_1\Big(\frac{M_{Z^\prime}x}{m_N}\Big)\Big[\frac{\frac{x^2_\Phi}{4}(10x^2_H+13x^2_\Phi+16x_H x_\Phi)}{\frac{1}{12}(103x^2_H+148x^2_\Phi+172x_Hx_\Phi)+x^2_\Phi}\Big],
\label{Eq:a16}
\end{equation}
where $x= m_N/T$. Hence for $x\lesssim \frac{m_N}{M_{Z^\prime}}$, Eq.~\ref{Eq:a16} becomes,
\begin{equation}
<\sigma v> \simeq \frac{\pi}{64}g^{\prime2}\frac{M^2_{Z^\prime}}{m^4_N}x^4\Big[\frac{\frac{x^2_\Phi}{4}(10x^2_H+13x^2_\Phi+16x_H x_\Phi)}{\frac{1}{12}(103x^2_H +148x^2_\Phi+172x_H x_\Phi)+x^2_\Phi}\Big],
\label{Eq:a17}
\end{equation}
where we have used the approximation $K_1\Big(\frac{M_{Z^\prime}x}{m_N}\Big)\simeq \frac{m_N}{M_{Z^\prime}x}$ for $x\lesssim\frac{m_N}{M_{Z^\prime}}$. For $x\gtrsim \frac{m_N}{M_{Z^\prime}}$, the first order modified Bessel function is exponentially suppressed and hence, $<\sigma v> \simeq 0$. This is because in this limit, $T<M_{Z^\prime}$ and the energy of the SM particles are not enough to produce $Z^\prime$ resonance and hence DM creation rate drops. Now integrating the Boltzmann equation from $x=x_i$ to $x=\frac{m_N}{M_{Z^\prime}}$ and putting the initial condition $Y({x_i})=0$, we obtain analytically the comoving DM density $Y_{\infty}\simeq Y_{\mathrm{tot}}(x=\frac{m_N}{M_{Z^\prime}})$ as,
\begin{equation}
Y_{\infty}\simeq 4.2\times 10^{-6}g^{\prime 2}\Big(\frac{M_{pl}}{M_{Z^\prime}}\Big)\Big[\frac{x^2_\phi(10x^2_h+13x^2_\phi+16x_Hx_\Phi)}{\frac{1}{12}(103x^2_H+148x^2_\phi+172x_Hx_\Phi)+x^2_\Phi}\Big].
\end{equation}
Hence, the relic density of DM becomes,
\begin{equation}
\begin{split}
\Omega_Nh^2 \simeq 0.12 \Big(\frac{g^\prime}{6.54\times 10^{-9}}\Big)^2 \Big(\frac{m_N}{10~\textrm{keV}}\Big)\Big(\frac{10~\textrm{GeV}}{M_{Z^\prime}}\Big)\Big[\frac{x^2_\Phi(10x^2_H+13x^2_\Phi+16x_H x_\Phi)}{\frac{1}{12}(103x^2_H+148x^2_\Phi+172x_H x_\Phi)+x^2_\Phi}\Big]. 
\end{split}
\label{comega}
\end{equation}
On the other hand, if $Z^\prime$ can decay to top quark, then the relic density of the DM in $T_R> M_Z^\prime> m_N$ limit is modified as,
\begin{equation}
\begin{split}
\Omega_Nh^2\simeq 0.12 \Big(\frac{g^\prime}{6.54\times 10^{-9}}\Big)^2 \Big(\frac{m_N}{10~\textrm{keV}}\Big)\Big(\frac{10~\textrm{GeV}}{M_{Z^\prime}}\Big)\Big[\frac{x^2_\Phi(10x^2_H+13x^2_\Phi+16x_Hx_\Phi)}{(10x^2_H+13x^2_\Phi+16x_Hx_\Phi)+x^2_\Phi}\Big].
\end{split}
\label{4.30}
\end{equation}
Thus the DM relic density increases with $m_N$ and decreases with $M_{Z^\prime}$. The relic density also increases with increasing $U(1)^\prime$ gauge coupling. As in the previous cases, two different values of $x_\Phi$ can give the same relic density for a given $x_H$.
\begin{figure}
\centering
\begin{subfigure}{0.51\textwidth}
    \includegraphics[width=\textwidth]{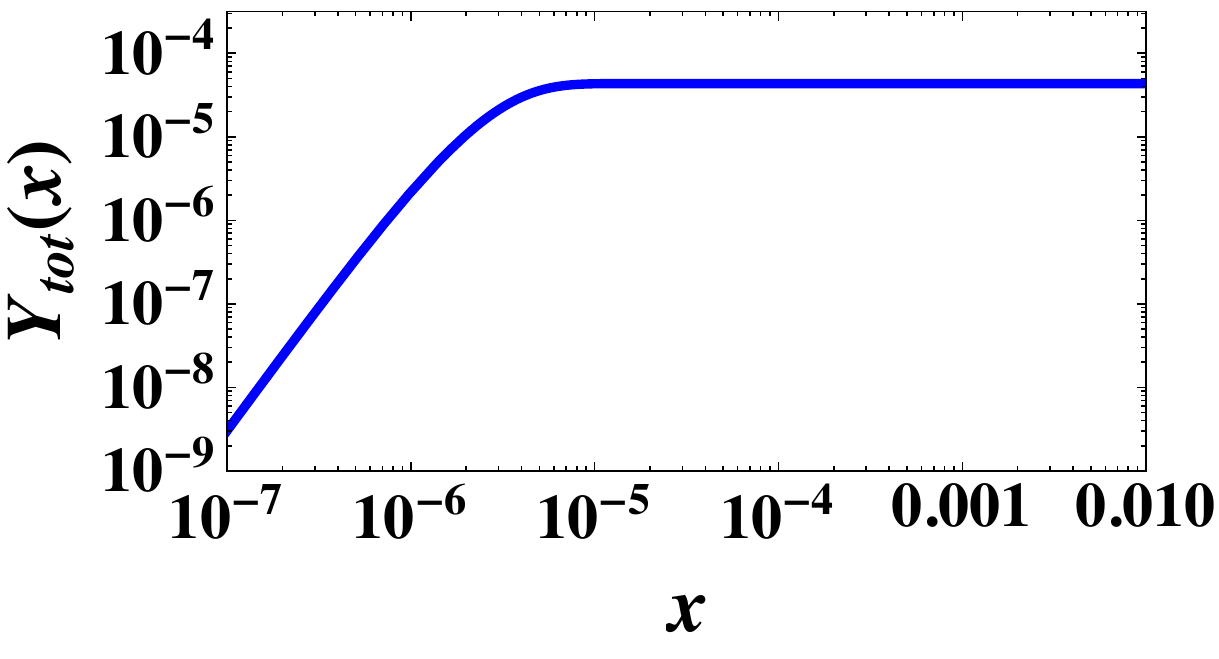}
    \caption{$Y(x)$ vs. $x$}
    \label{fig:intermediatefirst}
\end{subfigure}
\begin{subfigure}{0.48\textwidth}
    \includegraphics[width=\textwidth]{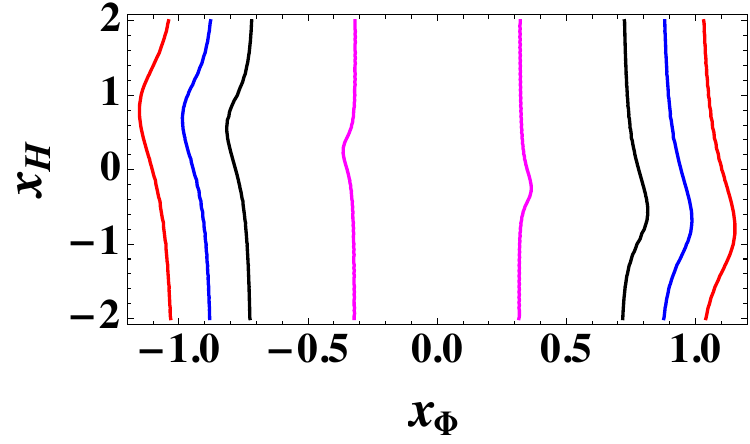}
    \caption{ $x_H$ vs. $x_\Phi$  }
    \label{fig:intermediatesecond}
\end{subfigure}
\caption{(a) DM yield $Y(x)$ vs. $x=\frac{m_N}{T}$ for case C ($T_R> M_Z^\prime> m_N$). (b) Contour plots of $x_H$ vs. $x_\Phi$ that satisfy the correct relic density $\Omega_N h^2=0.12$ for Case C.}
\label{Fig:mkk3}
\end{figure}

We obtain the DM yield as a function of $x$ in FIG.~\ref{fig:intermediatefirst} for case-C by numerically solving the Boltzmann equation as before. We have fixed $M_{Z^\prime}=10~\rm{GeV}$, $m_N=10~\rm{keV}$, $g^\prime=2\times 10^{-9}$ and $x_\Phi=1$, $x_H=-1.2$ in plotting this figure. The resultant $Y_\infty=4.36\times10^{-5}$ gives the correct relic density. Note that here, the freeze-in occurs at $x \sim m_N/M_{Z^\prime}$ (i.e., at $T\simeq M_{Z^\prime}$).

In FIG.~\ref{fig:intermediatesecond} we obtain the contour lines along which the relic density is satisfied. The dependence of relic density on $x_H$ is weak however, the relic density strongly depends on $x_\Phi$. The red, blue, magenta and black lines correspond to $(M_{Z^\prime}, g^\prime, m_N)=$ (10 GeV, $6\times 10^{-9}$, 10 keV), (10 GeV, $7\times 10^{-9}$, 10 keV), (10 GeV, $6\times 10^{-9}$, 100 keV), (5 GeV, $6\times 10^{-9}$, 10 keV).

For $M_{Z^\prime}=10~\rm{GeV}$ and $m_N=10~\rm{keV}$, the value of the  $B-L$ gauge coupling that gives the correct relic density is given as, $g_{B-L}=6.62\times 10^{-9}$. The $U(1)^\prime$ coupling and $U(1)_{B-L}$ coupling are related as,
\begin{equation}
g^\prime = 0.98 \times g_{B-L}
\Big[\frac{x^2_\phi(10x^2_H+13x^2_\Phi+16x_Hx_\Phi)}{\frac{1}{12}(103x^2_H+148x^2_\Phi+172x_Hx_\Phi)+x^2_\Phi}\Big]^{-\frac{1}{2}},
\label{d}
\end{equation}
by satisfying the relic density constraints. In the limit of $x_H\gg x_\Phi$ and $x_\Phi=1$, Eq. \ref{d} becomes $g^\prime=6.01\times 10^{-12}{\sqrt{\frac{M_{Z^\prime}}{m_N}}}$. Hence, in this limit, $g^\prime$ does not depend on $x_H$ unlike the cases A and B.
\section{Bounds on $Z'$ Mass from high energy, intensity and lifetime frontier Experiments}\label{sec5}
The $Z^\prime$ in the $U(1)^\prime$ model can be searched for in various experiments. In particular, the ATLAS experiment \cite{ATLAS:2019erb} can probe a heavy $Z^\prime$ boson ($M_{Z^\prime} \sim \mathcal{O}(\mathrm{TeV}))$ whereas a light $Z^\prime$ ($M_{Z^\prime} \sim 10^{-2}~\mathrm{GeV}-5~\mathrm{GeV})$ can be probed in DUNE \cite{Dev:2021qjj} and the future lifetime frontier experiments such as FASERs~\cite{Ariga:2018uku,Ariga:2019ufm} and ILC beam dump~\cite{Asai:2020xnz,Asai:2021ehn}. For the cases A, B, and C, we compare the parameter space allowed by the DM relic density with the existing astrophysical and collider constraints as well as with the parameter space that can be probed in the future lifetime frontier experiments, for different values of the scalar charges.
\subsection{Bounds on heavy $Z^\prime$}
The ATLAS experiment has given bounds on the cross section for the process $pp\rightarrow Z^\prime\rightarrow 2e, ~2\mu$ for the Sequential Standard Model (SSM) \cite{Altarelli:1989ff}. We have used this to obtain the bounds in the $M_{Z^\prime}-g^\prime$ planes for models with different values of $x_H$ and $x_\Phi$. The bound on the $g^\prime$ value corresponding to a given $M_{Z^\prime}$ value is then obtained as, 
\begin{equation}
g^\prime=\sqrt{{{g^\prime}^2_{\mathrm{Model}}}\frac{\sigma^{\mathrm{Observed}}_{\mathrm{ATLAS}}}{\sigma_{\mathrm{Model}}}},
\end{equation}
where $\sigma^{\mathrm{Observed}}_{\mathrm{ATLAS}}$ is the observed bound from ATLAS and $\sigma_{\mathrm{Model}}$ is the calculated value of the cross section for the given model. 
\begin{figure}
\centering
\includegraphics[width=7cm]{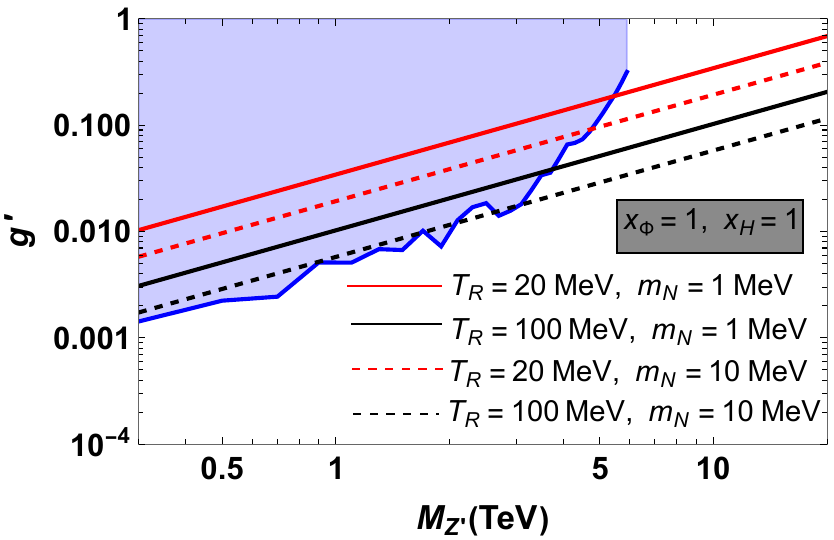}
\includegraphics[width=7cm]{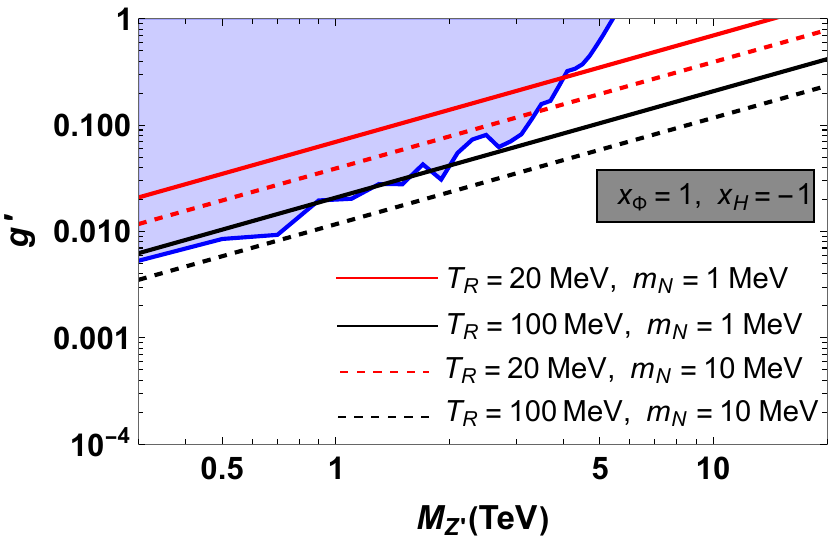}
\includegraphics[width=7cm]{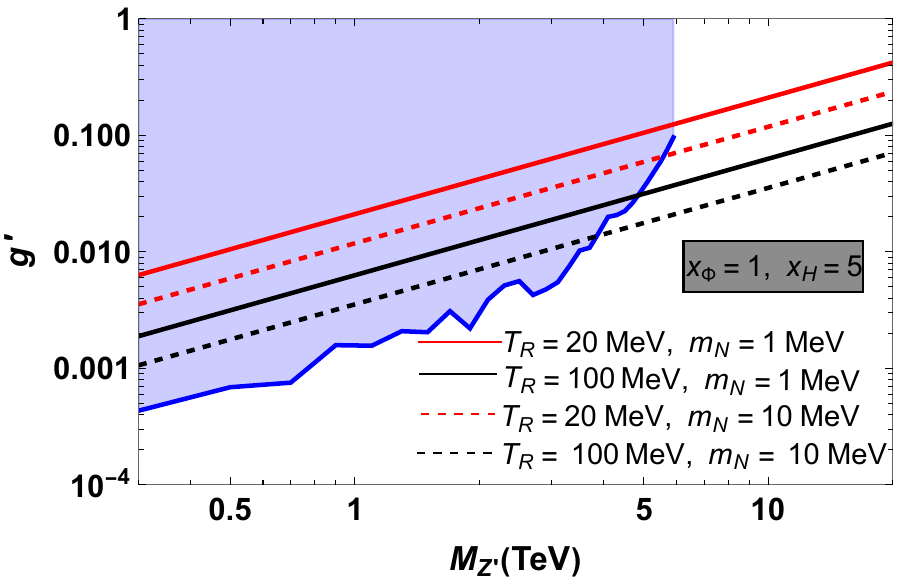}
\includegraphics[width=7cm]{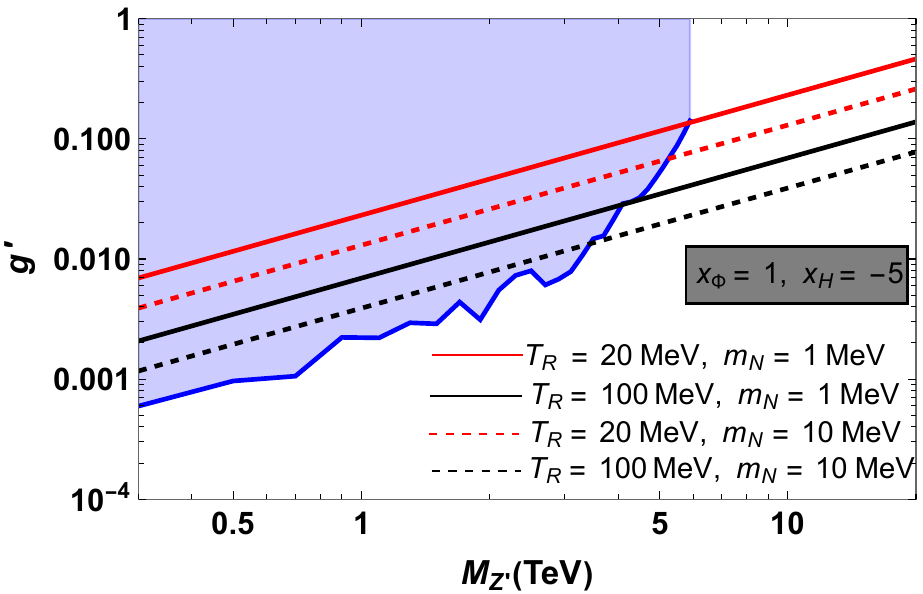}
\caption{The parameter space for the DM relic abundance in the $g^\prime-M_{Z^\prime}$ plane for different combinations of the scalar charges $x_H=1$, $-1$, $5$ and $-5$ for $x_\Phi=1$ in the $U(1)^\prime$ model, for the heavy $Z^\prime$ case $(M_{Z^\prime}> T_R> m_N)$. }
\label{fig:high}
\end{figure}
In FIG. \ref{fig:high} we have shown the variation of $g^\prime$ with $M_{Z^\prime}$ for different values of $x_H$, $x_\Phi$, $T_R$ and $m_N$. The heavy $Z^\prime$ considered here corresponds to the case-A of DM production discussed in Section \ref{subsec4}. The solid red, dashed red, solid black and dashed black lines correspond to the lines that give the correct relic density $\Omega_N h^2=0.12$ for $(T_R, m_N)=(20~\mathrm{MeV}, 1~\mathrm{MeV})$, $(20~\mathrm{MeV}, 10~\mathrm{MeV})$, $(100~\mathrm{MeV}, 1~\mathrm{MeV})$ and $(100~\mathrm{MeV}, 10~\mathrm{MeV})$ respectively. The blue shaded regions are disfavoured by the ATLAS search for a heavy $Z^\prime$ \cite{ATLAS:2019erb}. The parts of the red and black lines in the white region correspond to the points of the parameter space that give the correct relic density for a given $T_R$ and $m_N$ and are consistent with existing collider bounds. These regions can be probed in the future runs of the collider experiments. The value of the gauge coupling that gives the correct relic density increases with increasing $M_{Z^\prime}$, as we have already seen in Section \ref{subsec4}. Also we obtain stringent bounds on $g^\prime$ for larger values of $m_N$ and $T_R$.
\subsection{Bounds on light $Z^\prime$}
To obtain the correct relic density of the sterile neutrino DM through light $Z^\prime$ portal, the coupling should be very small as we have seen before. The parameter space for such a light $Z^\prime$ can be probed in several proposed proton beam dump experiments like FASERs \cite{FASER:2018eoc}, DUNE \cite{Dev:2021qjj} and electron/positron beam dump experiment like ILC-BD \cite{Sakaki:2020mqb,Asai:2021xtg}. We also compare our constraints on $U(1)^\prime$ gauge coupling with sensitivities for several existing electron beam dump experiments such as Orsay \cite{Davier:1989wz}, E137 \cite{Bjorken:1988as}, E141 \cite{Riordan:1987aw}, KEK \cite{Konaka:1986cb}, and proton beam dump experiments such as LSND \cite{LSND:1997vqj}, CHARM \cite{CHARM:1985anb}, $\nu-\textrm{Cal}$ \cite{Blumlein:2011mv,Blumlein:2013cua}, Nomad \cite{NOMAD:2001eyx}, PS191 \cite{Bernardi:1985ny}. We also compare our results with SN1987A (SFHo20.0 profile with $20.0M_\odot$ progenitor mass) \cite{Croon:2020lrf}. We obtain these constraints on $g^\prime-M_{Z^\prime}$ plane for different scalar charge combinations. 

\begin{figure}
\centering
\includegraphics[width=7cm]{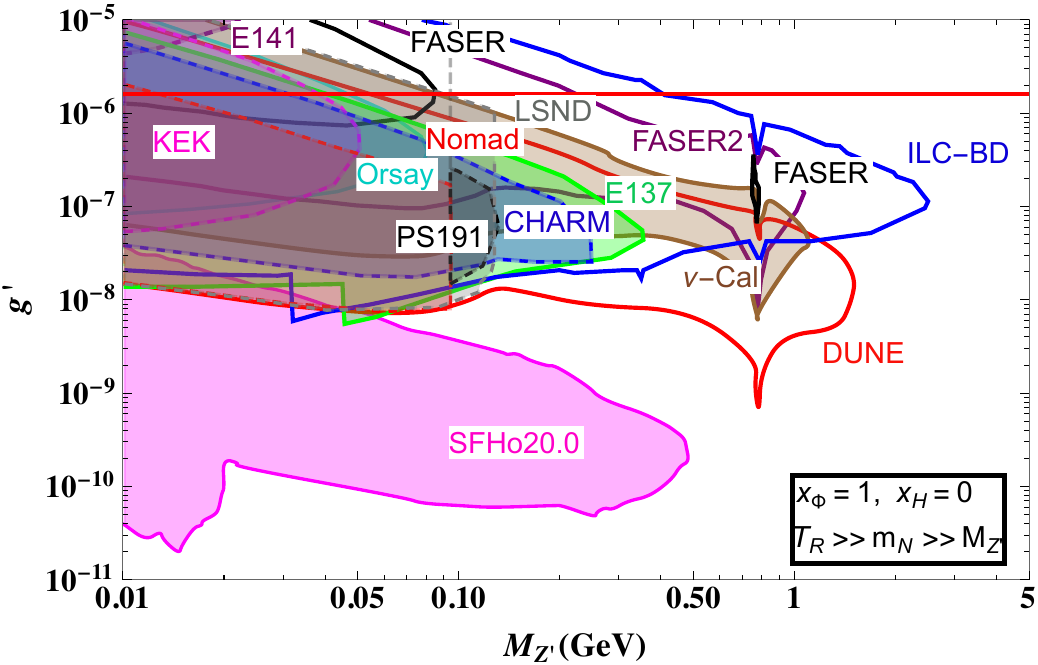}
\includegraphics[width=7cm]{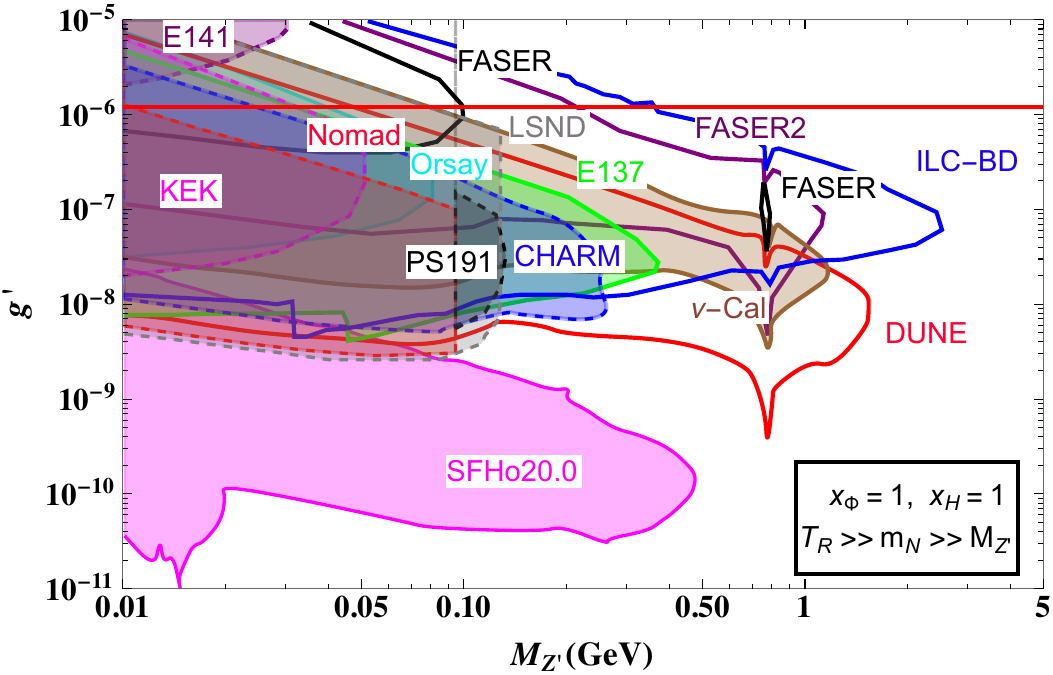}
\includegraphics[width=7cm]{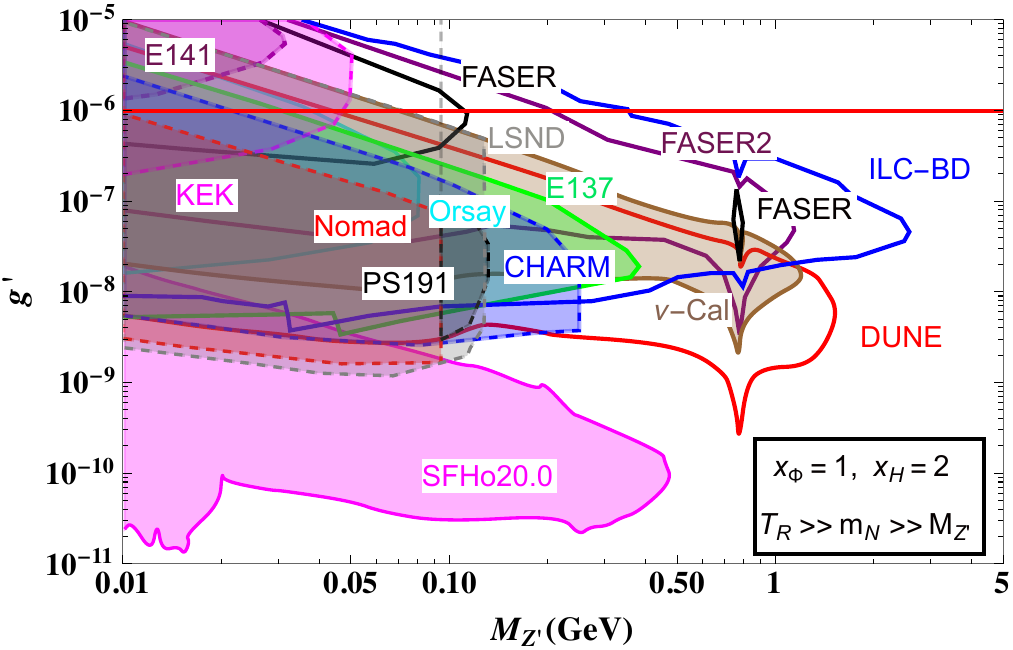}
\includegraphics[width=7cm]{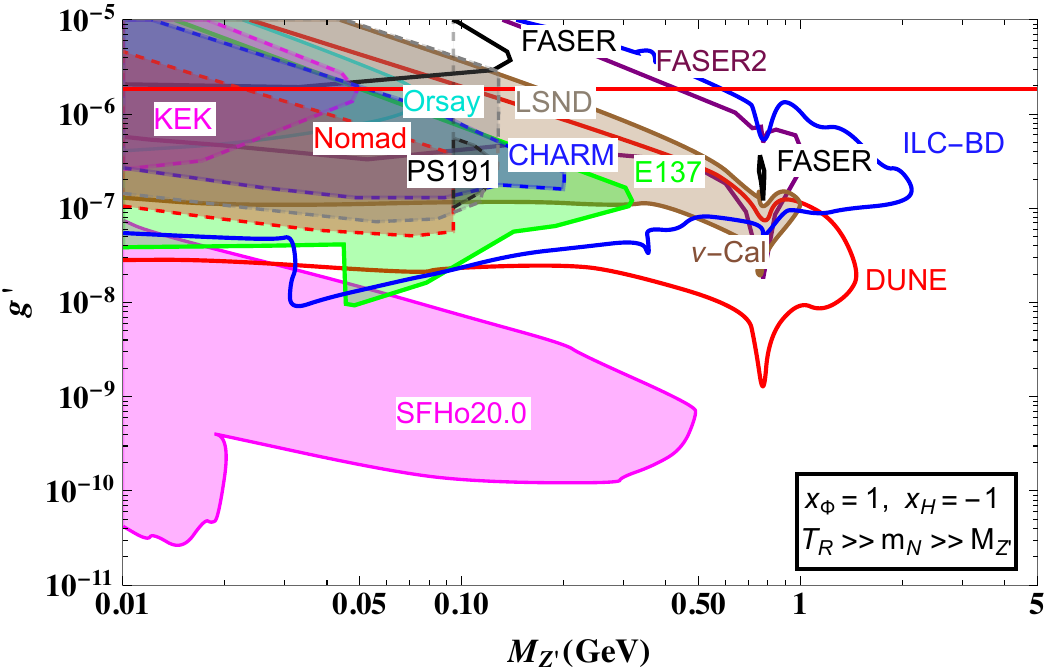}
\caption{The parameter space for the DM relic abundance in the $g^\prime-M_{Z^\prime}$ plane for different combinations of the scalar charges $x_H=0$, $1$, $2$ and $-1$ for $x_\Phi=1$ in the $U(1)^\prime$ model, for the case B $(T_R> m_N> M_{Z^\prime})$. }
\label{Fig:ml}
\end{figure}
\begin{figure}
\centering
\includegraphics[width=7cm]{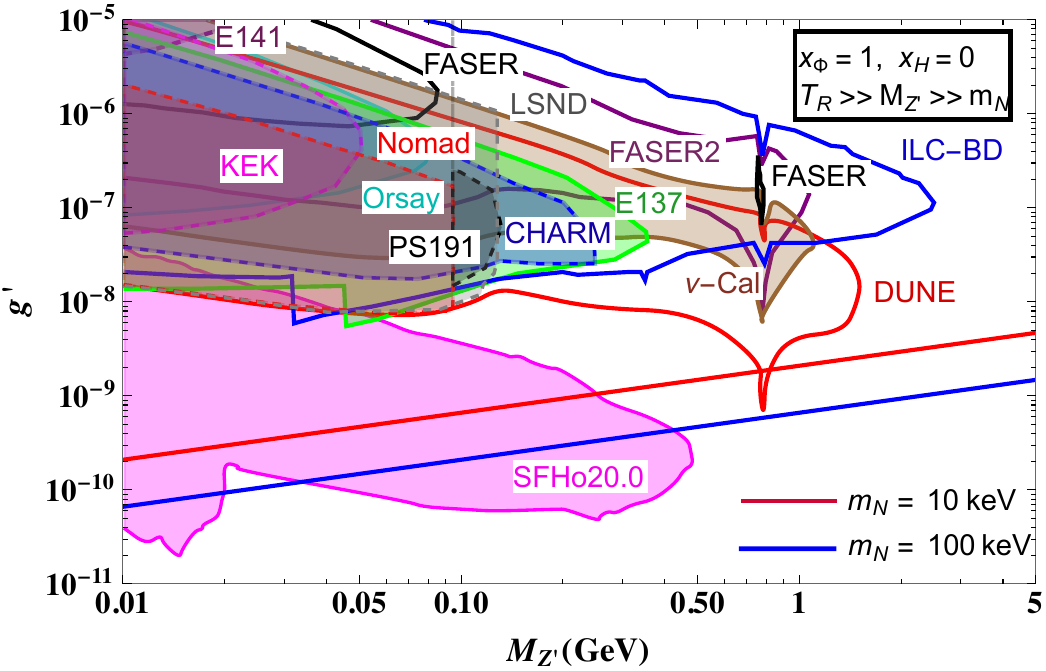}
\includegraphics[width=7cm]{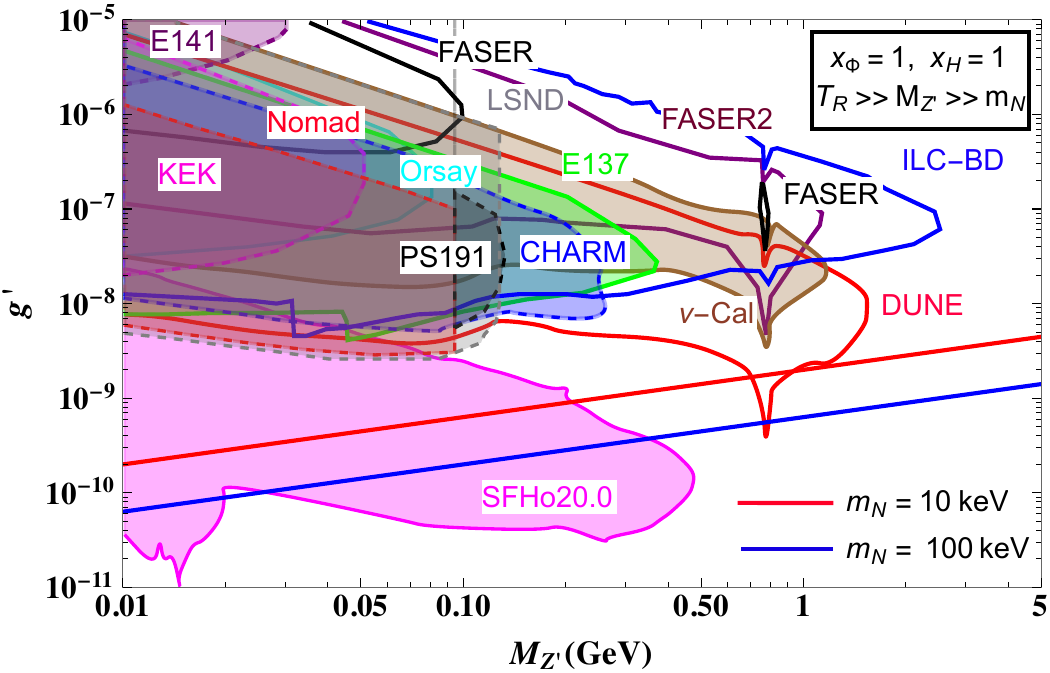}
\includegraphics[width=7cm]{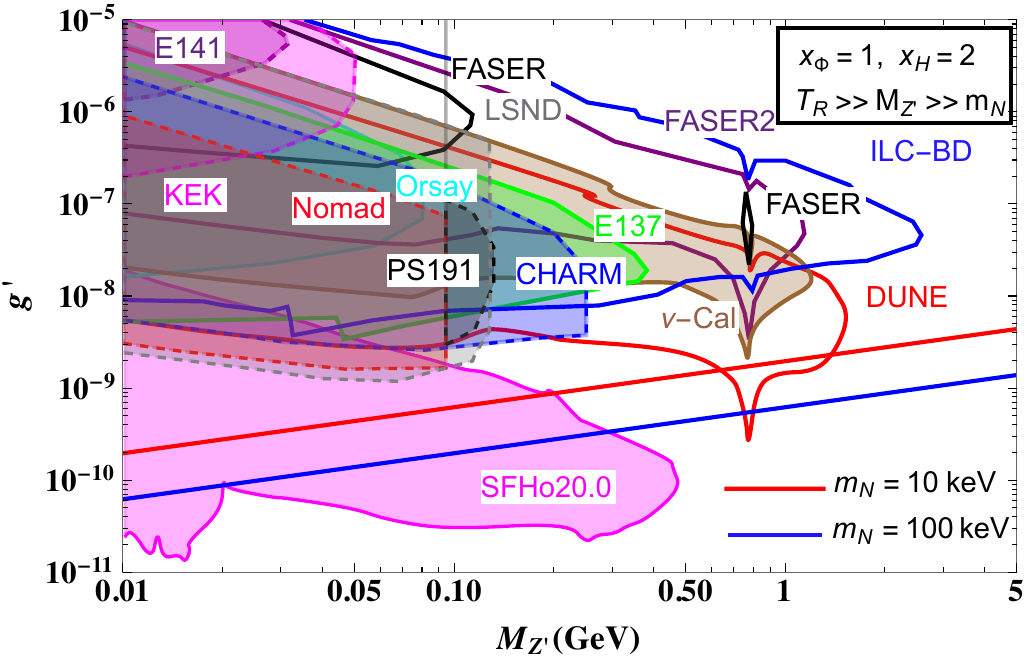}
\includegraphics[width=7cm]{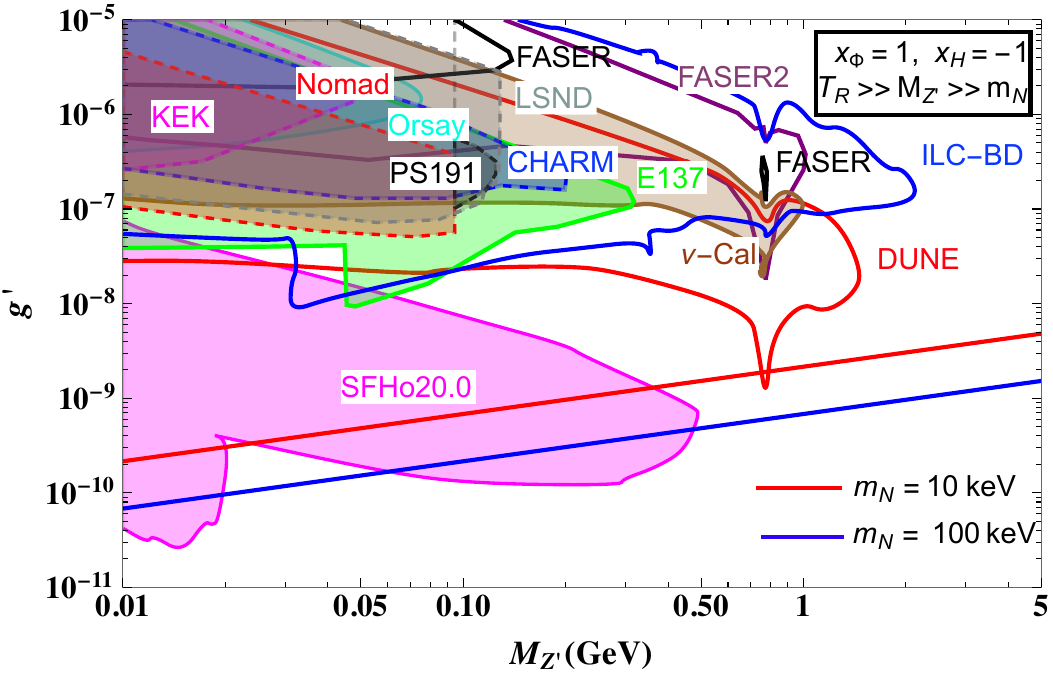}
\caption{The parameter space for the DM relic abundance in the $g^\prime-M_{Z^\prime}$ plane for different combinations of the scalar charges $x_H=0$, $1$, $2$ and $-1$ for $x_\Phi=1$ in the $U(1)^\prime$ model, for case C $(T_R> M_{Z^\prime}> m_N)$.}
\label{Fig:mlb}
\end{figure}

The parameter spaces that satisfy the correct DM relic abundance in the $g^\prime-M_{Z^\prime}$ planes for different values of $x_H$ and $x_\Phi$ are shown in FIGs. \ref{Fig:ml} and \ref{Fig:mlb}. These two figures correspond to the cases B and C of DM production discussed in Sections \ref{subsec5} and \ref{int} respectively. We use the constraints of different existing observations and experiments \cite{Asai:2022zxw} like SN1987A (magenta solid line), E137 (green solid line), E141 (purple dashed line), KEK (magenta dashed line), Nomad (red ashed line), CHARM (blue dashed line), $\nu-\textrm{Cal}$ (brown solid line), PS191 (black dashed line), Orsay (cyan solid line), LSND (gray dashed line), and several future experiments like FASER (black solid line), FASER2 (purple solid line), DUNE (red solid line), ILC-BD (blue solid line) for the different scalar charge combinations. The relic density of the sterile neutrino DM in FIG. \ref{Fig:ml} $(T_R> m_N> M_{Z^\prime})$ does not depend on the masses of the $Z^\prime$ as well as the DM candidate mass as discussed in Sec.~\ref{subsec5}. In FIG.~\ref{Fig:mlb}, we obtain the variation of $g^\prime$ with respect to $M_{Z^\prime}$ for $m_N=10~\mathrm{keV}$ (red line), $100~\mathrm{keV}$ (blue line) satisfying the correct DM relic density for case C ($T_R>M_{Z^\prime}>m_N$). The gauge coupling increases with $M_{Z^\prime}$ and decreases with $m_N$ as discussed in Sec.~\ref{int}. Some of the parameter space of $Z^\prime$, giving the correct relic density can be probed at the lifetime frontier experiments in the near future.

FIG.~\ref{Fig:ml} and FIG.~\ref{Fig:mlb} can distinguish the search for light $Z^\prime$ (cases B and C) in the $U(1)_{B-L}$ model and general $U(1)^\prime$ model. With increasing $x_H$ and $x_\Phi$, the $g^\prime$ can be decreased to have a correct relic density. Though the dependence of relic density on $x_H$ for the intermediate $Z^\prime$ case is very weak.

For cases B and case C, we could have taken the reheating temperature $T_R$ of the order of $\mathrm{GeV}$ scale as long as $(T_R>m_N>M_{Z^\prime})$ and $(T_R>M_{Z^\prime}>m_N)$ are satisfied, respectively. However, from phenomenological point of view, we are trying to search for $Z^\prime$ in lifetime frontier experiments such as DUNE, FASER, ILC-BD and these experiments are mostly sensitive for $M_{Z^\prime}\sim \mathcal{O}(\mathrm{GeV})$. Therefore, with the choice of $T_R>\mathcal{O}(\mathrm{TeV})$, we have a larger parameter space for $Z^\prime$ to be probed.
\section{INTEGRAL Anomaly and Galactic 511 $\rm{keV}$ line}\label{inte}
In this section, we briefly discuss how this model can also serve as a solution to the galactic $511$ keV line observed by the INTEGRAL satellite~\cite{Knodlseder:2003sv,Jean:2003ci}, provided the DM mass is in the MeV range. The authors of \cite{Boehm:2003bt,Fayet:2004bw,Gunion:2005rw,Boehm:2006gu,Hooper:2008im,Hooper:2004qf,Picciotto:2004rp,Kawasaki:2005xj,Takahashi:2005kp,Kasuya:2005ay,Pospelov:2007xh,Kasuya:2006kj,Chun:2006ss} have pointed out that the decay or annihilation of an MeV scale sterile neutrino DM can produce low energy $e^+ e^-$ pairs which could be the origin of the signal. Due to the collision with background baryonic matter, the positrons loose kinetic energy and form positronium which is an unstable bound system of electron and positron.
Thereafter depending on their relative spin states, the positronium annihilates into photons of $511~\mathrm{keV}$ line. 
The DM particles involved in the process are very light, as light as the electron, because otherwise the positrons would be too energetic to produce positronium.
The $\gamma$ ray photon flux observed by the INTEGRAL satellite at $2\sigma$ level is~\cite{Jean:2003ci},
\begin{equation}
\Phi_{511}=(1.05\pm 0.06)\times 10^{-3} ~\rm{photon/cm^{2}/sec}.
\end{equation}
In our model, the leptonic charged current Lagrangian can be written as,
\begin{equation}
\mathcal{L}_{cc}=\frac{g}{\sqrt{2}}W_\mu \bar{l}_l\gamma^\mu P_L(U_{li}\nu_i+V_{lj}N_j)+\rm{h.c.},
\end{equation}
where $P_L$ is the left chiral projection operator and $g$ is the $SU(2)_L$ gauge coupling. We can also write the neutral current Lagrangian for the neutrinos as,
\begin{equation}
\begin{split}
\mathcal{L}_{NC}=-\Big(\frac{g}{2\cos\theta_W}Z_\mu+g^\prime x_\nu {Z^\prime}_\mu\Big)\Big[(U^\dagger U)_{ij}\bar{\nu_i}\gamma^\mu P_L\nu_j+(U^\dagger V)_{ij}\bar{\nu_i}\gamma^\mu P_LN_j+\\
(V^\dagger V)_{ij}\bar{N_i}\gamma^\mu P_LN_j\Big]
+h.c.,
\end{split}
\end{equation}
where $\theta_W$ is the Weinberg angle, $x_\nu$ is the $U(1)^\prime$ charge of neutrino and $g^\prime$ is the $U(1)^\prime$ gauge coupling. From the above Lagrangian, the off-shell $W$ mediated leptonic partial decay width of the sterile neutrino DM can be calculated as,
\bea
\Gamma^{W^\ast}(N_i\rightarrow \ell_{L_{\alpha}} \ell_{L_\beta}\nu_k)=\frac{G^2_F}{192\pi^3}m^5_{N_i}|V_{\alpha i}|^2|\rm{}U_{\rm{PMNS}\beta k}|^2.
\eea
Similarly the off-shell $Z$ mediated leptonic partial decay width of the sterile neutrino is given as,
\bea
\Gamma^{Z^\ast}(N_i\rightarrow \nu_\alpha \ell_{R_\beta} \ell_{R_k})=\frac{G^2_F}{192\pi^3}m^5_{N_i}|V_{\alpha i}|^2\delta_{\beta k}\sin^4\theta_W,
\eea
for the right handed fermions and,
\bea
\Gamma^{Z^\ast}(N_i\rightarrow \nu_\alpha \ell_{L_\beta} \ell_{L_k})=\frac{G^2_F}{192\pi^3}m^5_{N_i}|V_{\alpha i}|^2\delta_{\beta k}\frac{\cos^2 2\theta_W}{4},
\eea
for the left handed fermions, respectively. The contribution to the decay from the interference between the off-shell $W$ and off-shell $Z$ mediated modes can be written as, 
\bea
\Gamma^{Z^\ast/W^\ast}(N_i\rightarrow \nu_\alpha l_\alpha l_\alpha)=\frac{G^2_F}{96\pi^3}m^5_{N_i}|V_{\alpha i}|^2 \textrm{Re}[\rm{U}_{\rm{PMNS}_{ii}^{}}].
\eea
In our case, $`l'$ stands for electron and positron. Adding all the $W$ and $Z$ mediated partial decay widths, we obtain the total decay width relevant for our analysis as,
\begin{equation}
\Gamma(N\rightarrow \nu e^+ e^-)=5.278\times \frac{G^2_F}{192\pi^3}m^5_{N_i}|V_{\alpha i}|^2.
\label{Eq:s1}
\end{equation}
We consider $m_e\ll m_N$ in calculating the decay width. The $Z^\prime$ contribution is not taken into account because its effect is smaller than the $W$ and $Z$ mediated processes. The decay of pseudo-Dirac DM mediated by the off-shell heavy $Z^\prime$ in the $\mathrm{TeV}$ scale is negligibly small due to $Z^\prime$ mass suppression. The coupling between $Z^\prime$, DM and light active neutrino is proportional to the $\nu-\mathrm{DM}$ mixing and hence suppressed. Here, $G_F$ is the Fermi constant and $V_{\alpha i}$ includes the mixing angle. 
The 511 keV gamma ray flux from the decaying DM is given as~\cite{Hooper:2004qf,Picciotto:2004rp},
\begin{equation}
\Phi_{511}\sim 10^{-3}\Big(\frac{10^{27}\sec}{\Gamma(N\rightarrow \nu e^+ e^-)^{-1}}\Big)\Big(\frac{1~\textrm{MeV}}{m_N}\Big)~\rm{photon/cm^{2}/sec}.
\label{Eq:s2}
\end{equation}
From Eq.~\ref{Eq:s1} and Eq.~\ref{Eq:s2} we can find out the mixing angle for which the decaying DM can give rise to $511$ keV excess and, 
\begin{equation}
|V_{\alpha i}|^2\Big(\frac{m_N}{1~\textrm{MeV}}\Big)^4\simeq 5.44\times 10^{-24}.
\end{equation}
 For $1$ MeV DM the corresponding mixing angle is $|V_{\alpha i}|^2\sim 5.44\times 10^{-24}$. 
\begin{figure}
\centering
\includegraphics[width=12cm]{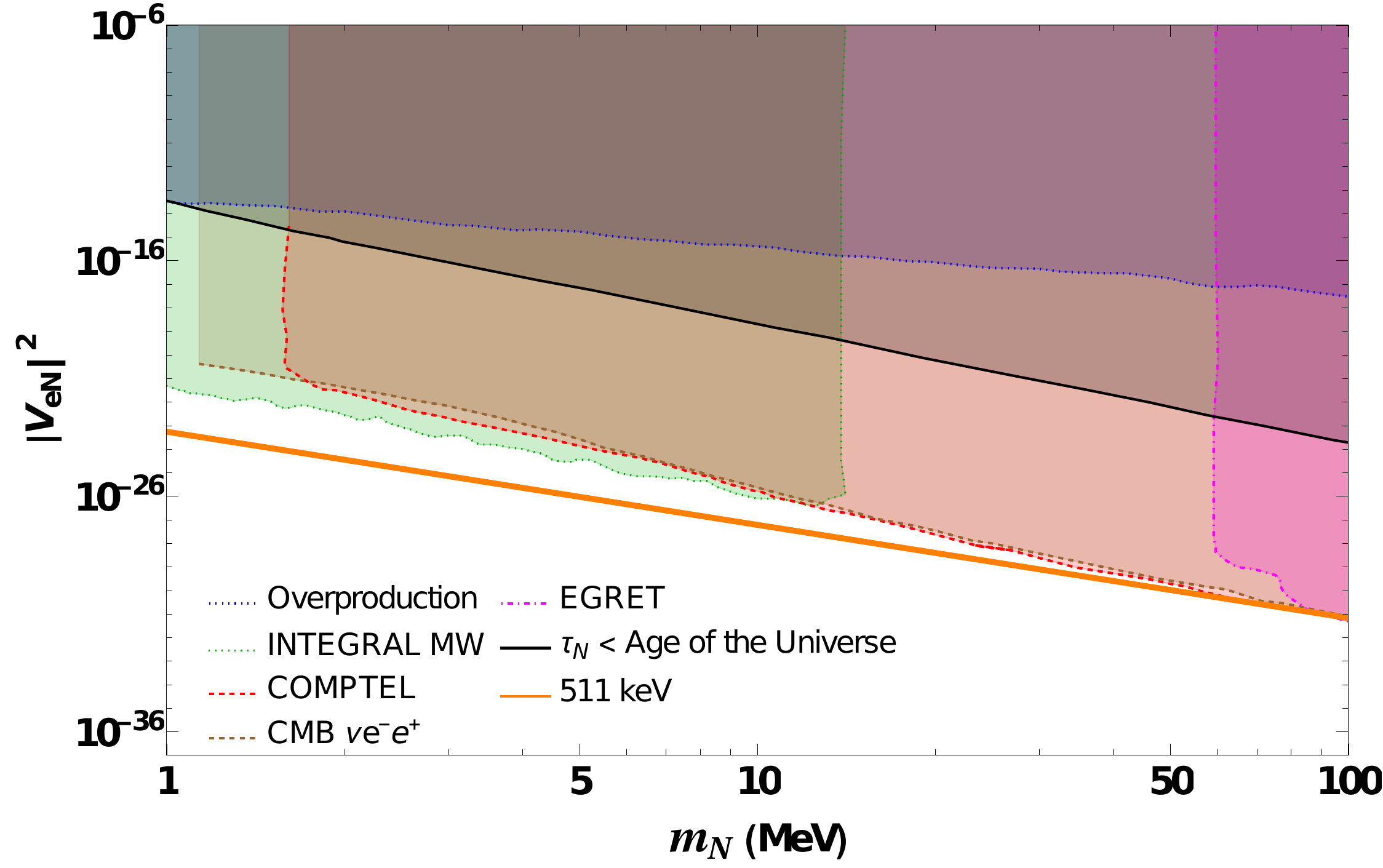}
\caption{The orange line corresponds to the values of the masses and mixing of the sterile neutrino DM that can produce the 511 keV line. We have also displayed the existing experimental bounds that were discussed in \cite{DeRomeri:2020wng}.}
\label{mix2}
\end{figure}
In Fig.~\ref{mix2}, the orange line corresponds to the values of the masses and mixing of the sterile neutrino DM that can produce the 511 keV line. We have also displayed the existing experimental constraints that were discussed in \cite{DeRomeri:2020wng}. The shaded regions are excluded. As can be seen from this figure, for 1 MeV $ \leq  m_N \leq$ 70 MeV, the values of the active-sterile mixing needed to produce the 511 keV line are consistent with all the existing bounds.
\section{Conclusions}
\label{sec6}
In this paper, we have investigated a class of models where the SM is augmented by an additional $U(1)^\prime$ gauge group, to offer explanations for two key phenomena: the observed relic density of DM through the freeze-in mechanism and the generation of light neutrino masses using the inverse seesaw mechanism. To keep the analysis general, we treat the charges of the two scalar particles (the SM Higgs and the $U(1)^\prime$ scalar) as independent parameters and express charges of all the fermions in terms of these scalar charges. In addition to the SM fermions, our model includes three right-handed neutrinos and three additional singlet fermions. The active light neutrinos get their mass through the inverse seesaw mechanism. The model involves two pairs of pseudo-Dirac neutrinos with masses around the TeV scale, contributing to the light neutrino masses and resulting in an almost massless lightest active neutrino. The other pair of pseudo-Dirac neutrinos serve as DM candidates in our model. The DM production occurs through the scattering of SM particles via the $Z^\prime$ portal. The reheating temperature at which DM is produced can vary in different scenarios, depending on the masses of the $Z^\prime$ boson and the DM particle.

We derive constraints on the scalar charges, denoted as $x_H$ and $x_\Phi$, by ensuring that the model satisfies the observed relic abundance of DM. We also establish constraints on the $U(1)^\prime$ gauge couplings in different mass ranges of the $Z^\prime$ boson. In the case of a heavy $Z^\prime$ boson, we determine constraints on the gauge coupling $g^\prime$ for various reheating temperatures and DM masses, and compare them with the constraints from the ATLAS experiment. For scenarios with light $Z^\prime$, we identify parameter spaces that can be tested in experiments such as DUNE, as well as future lifetime frontier experiments like FASERs, and ILC beam dump. We derive constraints on the $Z^\prime$ gauge boson and DM masses for various scalar charges $(x_H, x_\Phi)$ in our model. These constraints are distinct from those in the $U(1)_\textrm{B-L}$ model, emphasizing the unique features of our analysis. We have also pointed out that that when the mass of the pseudo-Dirac DM is $\gtrsim 1~\mathrm{MeV}$ and the $Z^\prime$ particle is heavy, the decay of the DM into active neutrinos could offer a solution to the enduring mystery of the 511 keV gamma-ray line detected by the INTEGRAL satellite in our galaxy.
\section*{Acknowledgements} S.G. acknowledges the J.C Bose Fellowship (JCB/2020/000011) of Science and Engineering Research Board of Department of Science and Technology, Government of India. She also acknowledges Northwestern University for hospitality and the United States-India Education Foundation (USIEF) for supporting her stay through the Fullbright-Nehru Professional and Academic Excellence Fellowship. 

\bibliography{bibliography}
\bibliographystyle{utphys}
\end{document}